\documentclass[apj]{emulateapj}
\usepackage{apjfonts}
\usepackage{epsfig}
\usepackage{amsmath}
\usepackage{amssymb}
\usepackage{amsfonts}
\usepackage{lscape}
\usepackage{natbib}          
\shorttitle{DEEP OBSERVATIONS OF LMXBS IN NGC~4697}
\shortauthors{SIVAKOFF ET AL.}

\defcitealias{JCF+2004}{J2004}
\defcitealias{BIS+2006}{B2006}
\defcitealias{SJS+2007}{S2007}

\defcitealias{ISB2000}{I2000}
\defcitealias{SIB2000}{Paper~I}
\defcitealias{SIB2001}{Paper~II}
\defcitealias{SSJ2005}{Paper~III}
\defcitealias{SJJ+2008a}{Paper~IV} 
\defcitealias{SJJ+2008b}{Paper~V}

\newcommand\cennodata{\multicolumn{1}{c}{\nodata}}

\begin{document}

\title{Deep Chandra X-ray Observations of Low Mass X-ray Binary Candidates
in the Early-Type Galaxy NGC~4697}
\author{
Gregory R. Sivakoff\altaffilmark{1,2},
Andr\'{e}s Jord\'{a}n\altaffilmark{3,4},
Adrienne M. Juett\altaffilmark{5},
Craig L. Sarazin\altaffilmark{1},
Jimmy A. Irwin\altaffilmark{6}
}

\altaffiltext{1}{
Department of Astronomy,
University of Virginia,
P. O. Box 400325,
Charlottesville, VA 22904-4325, USA;
sarazin@virginia.edu}
\altaffiltext{2}{
Current Address:
Department of Astronomy,
The Ohio State University,
4055 McPherson Laboratory
140 W. 18th Avenue, Columbus, OH 43210-1173, USA;
sivakoff@astronomy.ohio-state.edu}
\altaffiltext{3}{%
Clay Fellow,
Harvard-Smithsonian Center for Astrophysics,
60 Garden Street,
MS-67, Cambridge, MA 02138, USA;
ajordan@cfa.harvard.edu}
\altaffiltext{4}{%
Departamento de Astronom\'{\i}a y Astrof\'{\i}sica, 
Pontificia Universidad Cat\'olica de Chile,
Casilla 306, Santiago 22, Chile}
\altaffiltext{5}{%
NASA Postdoctoral Fellow,
Laboratory for X-ray Astrophysics,
NASA Goddard Space Flight Center,
Greenbelt, MD 20771, USA;
adrienne.m.juett@nasa.gov}
\altaffiltext{6}{
Department of Astronomy,
909 Dennison Building,
University of Michigan,
Ann Arbor, MI 48109-1042, USA;
jairwin@umich.edu}

\begin{abstract}
{\it Chandra} X-ray observations routinely resolve tens to hundreds of low-mass
X-ray binaries (LMXBs) per galaxy in nearby massive early-type galaxies. These
studies have raised important issues regarding the behavior of this population
of remnants of the once massive stars in early-type galaxies, namely the
connection between LMXBs and globular clusters (GCs) and the nature of the LMXB
luminosity function (LF). In this paper, we combine five epochs of {\it Chandra}
observations and one central field {\it Hubble Space Telescope} Advance Camera
for Surveys observation of NGC~4697, one of the nearest, optically luminous
elliptical (E6) galaxies, to probe the GC-LMXB connection and LMXB-LF down to a
detection/completeness limit of $0.6/1.4\times 10^{37} {\rm \, ergs \, s}^{-1}$. 
We detect 158 sources, present their luminosities and hardness ratios, and
associate 34 LMXBs with GCs. We confirm that GCs with higher encounter rates
($\Gamma_h$) and redder colors (higher metallicity $Z$) are more likely to
contain GCs, and find that the expected number of LMXBs per GC is proportional
to $\Gamma_{h}^{0.79^{+0.18}_{-0.15}} \, (Z/Z_\odot)^{0.50^{+0.20}_{-0.18}}$,
consistent with fainter X-ray sources in Galactic GCs and LMXBs in Virgo
early-type galaxies. Approximately $11\pm2/8\pm2$\% of GCs in NGC~4697 contain
an LMXB at the detection/completeness limit. We propose that the larger
proportion of metal-rich GCs in NGC~4697 compared to the Milky Way explains why
these fractions are much higher than those of the Milky Way at similar
luminosities. We confirm that a broken power-law is the best fit to the LMXB-LF,
although we cannot rule out a cutoff power-law, and argue that this raises the
possibility that there is no universal form for the LMXB-LF in early-type
galaxies. We find marginal evidence for different LFs of LMXBs in GCs and the
field  and different spectra of GC-LMXBs and Field-LMXBs.
\end{abstract}

\section{Introduction}
\label{sec:n4697x_intro}

Observations with the {\it Einstein Observatory} revealed that early-type
galaxies can be luminous X-ray sources \citep{FJT1985}. Prior to the launch of
the {\it Chandra X-ray Observatory}, observations of X-ray faint galaxies,
galaxies with relatively low X-ray-to-optical luminosity ratios, indicated the
presence of two distinct spectral components: a soft ($\sim 0.2 {\rm \, keV}$)
component \citep{FKT1994,P1994,KFM+1996} and a hard ($\sim 5$--$10 {\rm \,
keV}$) component \citep{MKA+1997}. The soft component was attributed to
hot interstellar gas and the hard component to low-mass X-ray binaries (LMXBs;
\citealt{KFT1992}).
Starting with the \textit{Chandra X-ray Observatory} observation of the X-ray
faint elliptical, \object{NGC 4697}
\citep[hereafter Papers I and II]{SIB2000,SIB2001},
this picture has been confirmed; the majority of X-ray emission in X-ray
faint early-type galaxies has been resolved into X-ray point sources, whose
properties are consistent with LMXBs. With its ability to resolve LMXBs and
accurately measure their positions, {\it Chandra} has raised at least two major
issues regarding the behavior of the LMXBs as a population, namely the
connection between LMXBs and globular clusters (GCs) and the nature of the LMXB
luminosity function (LF).

From {\it Chandra}, a high percentage $(\sim 20-70\%)$ of LMXBs have
positions coincident with globular clusters
\citep[GCs,][]{ALM2001,SKI+2003}; only $\sim 10\%$ of Milky Way LMXBs
are found in GCs. This raised the question as to where LMXBs in
early-type galaxies formed. One early suggestion was that all LMXBs in
early-type galaxies formed in GCs, with LMXBs observed to be in the
field of the galaxy (Field-LMXBs) having escaped from GCs through
supernovae kick velocities, stellar dynamical processes, or the
dissolution of the GC due to tidal effects \citep{WSK2002}. Later
analyses have suggested that a significant fraction of Field-LMXBs
were formed in situ \citep{J2005,I2005}, although there is evidence
that more Field-LMXBs in lenticular galaxies, as opposed to elliptical
galaxies, may have originated in GCs and later escaped into the
field \citep{I2005}. The extent to which LMXBs can be used to probe GC
formation and evolution, as well as LMXB formation, depends critically on
understanding the GC/LMXB connection. Given that the primary binary
formation scenario in GCs is thought to involve dynamical interactions,
as opposed to the predominantly primordial nature of binaries formed
in the field, these different populations should trace different LMXB
formation histories and may trace different star formation histories
in early-type galaxies.

{\it Chandra} has also explored the LF of LMXBs. At the bright end of
the LF of NGC~4697, a possible break near the Eddington
limit of a $1.4 \, M_{\sun}$ neutron star (NS) was found in
\citetalias{SIB2000}. This break was argued to be due to the presence
of two LMXB populations, a black hole (BH) population at the bright end, and a
predominantly NS population at the faint end. \citet{BD2004} argued that such a
break may be due to ultracompact binaries. Although
\citet{KF2003} argue that no break was required in NGC~4697 after
correcting for incompleteness, they do find that a break near
$5\times10^{38} {\rm \, ergs \, s}^{-1}$ gives an improved fit compared
to a single power-law for a uniformly selected, incompleteness
corrected sample of 14 early-type galaxies. At the faint end,
\citet{VG2005} found that the LF in \object[NGC 5128]{Cen A}
flattens significantly below $5\times10^{37} {\rm \, ergs \, s}^{-1}$ and
follows the $dN/dL \propto L^{-1}$ law in agreement with the behavior
found for LMXBs in the Milky Way and the bulge of M31. One
interpretation of this break in the Milky Way LF is that there are two
populations of short $(\lesssim 20 {\rm \, hr})$ period LMXBs: those
where magnetized stellar winds dominate mass transfer, and those where
gravitational radiation drives the accretion \citep{PK2005}.
Measuring and understanding the LF of LMXBs in early-type galaxies has
clear implications in our understanding of the number and type of binary
systems that make up the zoo of LMXBs.

As more LMXBs are detected and characterized, more examples of extragalactic
LMXB candidates with extreme behaviors, such as supersoft sources (SSs) and
ultraluminous X-ray sources (ULXs) are being identified by {\it Chandra} and
studied in
greater detail. The SSs have very soft X-ray spectra $(\lesssim 75 {\rm \, eV})$
similar to SSs in our Galaxy and M31 that are generally believed to be accreting
white dwarfs (WDs); however, the bolometric luminosities of extragalactic SSs
can exceed the Eddington luminosity for a Chandrasekhar mass WD. One hypothesis
is that these sources contain intermediate-mass ($\sim 10^2$--$10^3 \,
M_{\sun}$) accreting BHs \citep{SGS+2002}. A great deal of attention has been
spent on ULXs, which we define as $L_{X} > 2 \times 10^{39} {\rm \, ergs \,
s}^{-1}$ (0.3--10.0 keV). However, except possibly for LMXB candidates in
\object{NGC 720} \citep{JCB+2003},
\object{NGC 1399}
\citep{ALM2001}, \object{NGC 1600} \citep{SSC2004},
and NGC~4482 \citep{MKZ+2007},
ULXs are generally not found within old
stellar systems beyond the number expected from
unrelated foreground or background sources \citep{IBA2004}.

By stacking multi-epoch {\it Chandra} observations of an early-type
galaxy, more, fainter LMXB candidates can be studied, and brighter
LMXB candidates can be studied in greater detail.
In this Paper, we report on multi-epoch {\it Chandra} X-ray
observations of
NGC~4697, one of the nearest
\citep[$11.3 {\rm \, Mpc}$; see footnote 18 of][]{JCB+2005}
optically luminous ($M_B < -20$) elliptical (E6) galaxy.
(There is a weak disk; however, it not comparable to those seen in lenticular
galaxies \citep{PDI+1990}). 
We adopted the 2MASS Point Source
Catalog
\citep{SCS+2006}
position of R.A.\ $= 12^{\rm h} 48^{\rm m} 35\fs90$ and Dec.\
$= -5\arcdeg48\arcmin02\farcs6$ as the location of the center of
NGC~4697 and the Third Reference Catalogue of Bright Galaxies (RC3)
optical photometry values
for the effective radius
($r_{\rm eff}=72\farcs0$),
position angle ($PA=70\arcdeg$, measured from north to east), and ellipticity
($e=0.354$), which assumes a de Vaucouleurs profile \citep{VVC+1992}.
From these values, we calculate the optical photometry's effective
semi-major distance ($a_{\rm eff} = 89\farcs6$).
This galaxy lies $\sim 5 {\rm \, Mpc}$
in front of the bulk of the galaxies in the Virgo cluster, and
is $18\fdg7$ south of \object{M87}, the galaxy at the dynamical
 center of the Virgo cluster.
We also report on the GC/LMXB connection as determined from a joint
observation of the central region by the {\it Hubble Space Telescope}
Advance Camera for Surveys \citep[{\it HST}-ACS;][]{FBB+1998}.

In \S~\ref{sec:n4697x_obs}, we discuss the observations and data
reduction of NGC~4697. The X-ray image and the detection of X-ray sources
are discussed in \S~\ref{sec:n4697x_image} and
\S~\ref{sec:n4697x_detections}. In \S~\ref{sec:n4697x_opt_ids}, we discuss
the optical counterparts from ground-based and {\it HST} observations.
We examine the GC/LMXB connection in detail in \S~\ref{sec:n4697x_gclmxb}.
The analyses of luminosity, hardness ratios, and spectra are considered
in \S~\ref{sec:n4697x_src_lum}--\ref{sec:n4697x_src_spectra}.
Finally, we summarize our
conclusions in \S~\ref{sec:n4697x_conclusion}. 
We concentrate on the variability of the 
X-ray sources in our companion paper (Sivakoff et al.\ 2008b, hereafter Paper V).
Unless otherwise noted,
all errors refer to $1 \sigma $ confidence intervals, count rates are
in the $0.3$--$6 {\rm \, keV}$ band, and fluxes and luminosities
are in the $0.3$--$10 {\rm \, keV}$ band, with absorption effects removed.

\section{Observations and Data Reduction}
\label{sec:n4697x_obs}

\subsection{\itshape Chandra X-ray Observatory}
{\it Chandra} has observed NGC~4697 five times,
2000 January 15, 2003 December 26, 2004 January 06, February 02, and
August 18, using the ACIS detector for live exposures of 39260, 39920,
35683, 38103, and $40046\, {\rm s}$
(Observations
\dataset[ADS/Sa.CXO#obs/00784]{0784},
\dataset[ADS/Sa.CXO#obs/04727]{4727},
\dataset[ADS/Sa.CXO#obs/04728]{4728},
\dataset[ADS/Sa.CXO#obs/04729]{4729}, and
\dataset[ADS/Sa.CXO#obs/04730]{4730}).
Observation 0784 was operated at
$-110 ^{\circ} \,{\rm C}$ with a frame time of $3.2 {\rm \, s}$, and the
ACIS-23678 chips were telemetered and cleaned in Faint mode.
Observations 4727, 4728, 4729, 4730 (hereafter the Cycle-5
observations) were operated at $-120 ^{\circ} \,{\rm C}$ with frame times
of $3.1 {\rm \, s}$, and the ACIS-35678 chips were telemetered and
cleaned in Very-Faint mode. Since the X-ray point spread function
(PSF) increases with the distance between point sources and the optical
axis of {\it Chandra}, the Cycle-5 pointings were determined to
maximize field-of-view (FOV) while placing the galaxy center close to
the optical axis and away from node boundaries on the S3 chip. The
analysis in this Paper is based on data from the S3 chip alone,
although a number of serendipitous sources were seen on the other
chips. The elliptical area common to all five observations on the S3
chip ($a < 220\arcsec$) corresponds to 74\% of the integrated light
for a de Vaucouleurs profile. Known aspect offsets were applied to
each observation. Our analysis includes only events with ASCA grades
of 0, 2, 3, 4, and 6. Photon energies were determined using the gain
files acisD1999-09-16gainN0005.fits (Observation 0784) and
acisD2000-01-29gain\_ctiN0001.fits (Cycle-5 observations). In the
latter cases, we corrected for the time dependence of the gain and the
charge-transfer inefficiency. All five observations were corrected for
quantum efficiency (QE) degradation and had exposure maps determined
at $750 {\rm \, eV}$. We excluded bad pixels, bad columns, and
columns adjacent to bad columns or chip node boundaries.

Although {\it Chandra} is known to encounter periods of high background
(``background flares''), which especially affect the
backside-illuminated S1 and S3 chips\footnote{See
\url{http://cxc.harvard.edu/contrib/maxim/acisbg/}\label{ftn:4697_bkg}.},
the use of local backgrounds and small extraction regions in the point source
analysis mitigates the effect of flaring. To avoid periods with extreme flaring
we only included times where the blank-sky rate was less than three times the
expected blank-sky rate derived from calibrated blank-sky backgrounds. For
Observation 0784, the S3 chip itself, excluding regions of emission, was used as
the observed blank-sky and checked against Maxim Markevitch's
aciss\_B\_7\_bg\_evt\_271103.fits blank-sky
background\footnotemark[\ref{ftn:4697_bkg}] using $0.3$--$10.0 {\rm \, keV}$
count rates. For the Cycle-5 observations, the other back-illuminated chip,
S1, was available. Here, we used this chip, excluding
regions of emission, and compared to the blank-sky background in CALDB using
$2.5$--$6.0 {\rm \, keV}$ count rates. Minimal time was lost in all observations
due to the binning used to check the rates; more extensive time was lost in
Observation 4729 due to a large background flare. No periods of data dropout
were observed. Final flare-filtered live exposure times for the five
observations were 37174, 39919, 35601, 32038, and $40044 {\rm \, s}$.

We registered the Cycle-5 observations astrometry against the
Observation 0784 astrometry (see \S~\ref{sec:n4697x_detections}).
No absolute astrometric correction was necessary (see
\S~\ref{sec:n4697x_cat_ids}.). For imaging and source detection only,
we created a merged events file and exposure map with a live exposure
time of $184775 {\rm \, s}$. In this Paper and Paper V, we analyze the 158
sources detected from the merged events file.

All {\it Chandra} observations were analyzed using {\sc ciao 3.1}%
\footnote{See \url{http://asc.harvard.edu/ciao/}.} 
with {\sc caldb 2.28} and NASA's {\sc ftools 5.3}%
\footnote{See
\url{http://heasarc.gsfc.nasa.gov/docs/software/lheasoft/}%
\label{ftn:heasoft}.}. Source positions and extraction regions were
refined using ACIS Extract 3.34%
\footnote{See \url{http://www.astro.psu.edu/xray/docs/TARA/%
ae\_users\_guide.html}.}. All spectra
were fit using {\sc xspec}\footnotemark[\ref{ftn:heasoft}].

\subsection{\itshape Hubble Space Telescope}

We observed the center of NGC~4697 with the {\it Hubble Space
Telescope Advance Camera for Surveys} (HST-ACS), acquiring two $375
{\rm \, s}$ exposures in the F475W ($g$) band, two $560 {\rm \,
s}$ exposures in the F850LP ($z$) band, and one $90 {\rm \, s}$
F850LP exposure. Source detection and characterization were performed
similarly to \citet{JBP+2004}, leading to a list of globular clusters
(GCs) and other optical sources. 
We fit PSF convolved \citet{K1966c} models to all
detected sources using the code KINGPHOT described in
\citet{JCB+2005}.
In addition to best-fit magnitudes $g$ and $z$, the code returns the best-fit
half-light radius $r_h$ for each source. Details concerning the HST-ACS
observation, data analysis, and optical source properties are given in
Jord\'{a}n et al.\ (2008, in preparation).

\section{X-ray Image}
\label{sec:n4697x_image}

\begin{figure*}
\plotone{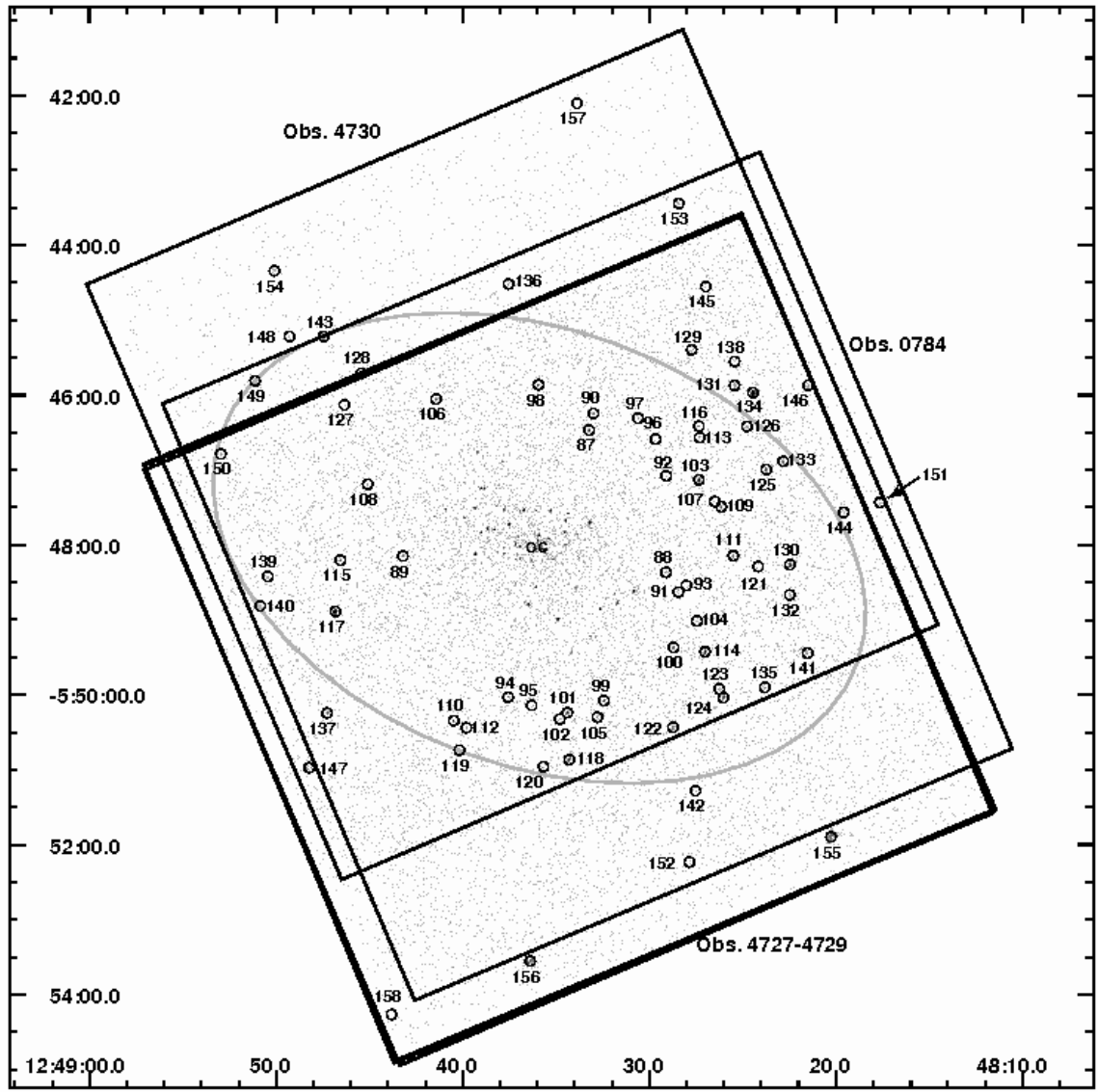}
\caption[Raw X-ray Greyscale X-ray Image of NGC~4697: S3 FOV]{
{\it Chandra} S3 image (0.3--$6 {\rm \, keV}$) of NGC~4697 from all
five observations combined.
This image has not been corrected for background or exposure, and has
not been smoothed.
The grey scale varies with the logarithm of the
X-ray surface brightness, which ranges from 1 to $25 {\rm \, count
\, pixel}^{-1}$. (The ACIS pixels are $0\farcs492$ square.) The
optical center of NGC~4697 is marked by ``OC''. The positions of
detected sources outside of the central $3\arcmin \times 3 \arcmin$ in
the image are indicated by their source numbers from
Table~\ref{tab:n4697x_src}; the source numbers are ordered by
increasing distance from the center of the galaxy.
The FOV of each observation is indicated by a labeled black square
and the D25 ellipse is shown.
(D25 is the elliptical isophote with a surface brightness of 25 mag
per arcsec in the $B$-band.)
\label{fig:n4697x_raw_whole}}
\end{figure*}

\begin{figure*}
\plotone{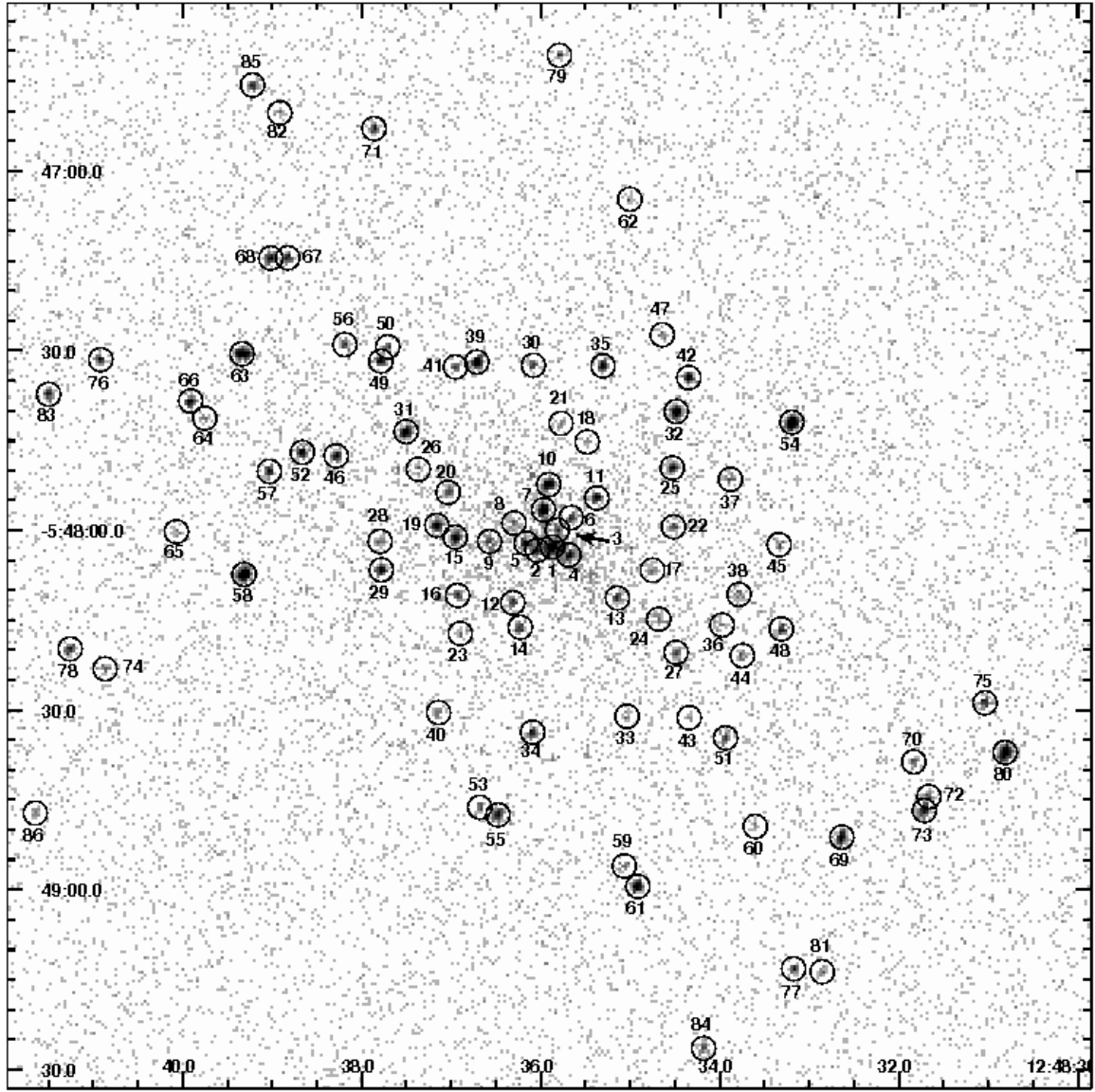}
\caption[Raw X-ray Greyscale Image of NGC~4697: Central $3\arcmin \times 3 \arcmin$ FOV]
{
{\it Chandra} S3 image (0.3--$6 {\rm \, keV}$) of the central
$3\arcmin \times 3 \arcmin$  of NGC~4697 from all five observations.
The optical center of NGC~4697 lies within the circle for Source 1.
The grey scale is the same as in Figure~\ref{fig:n4697x_raw_whole}.
The positions of detected sources in the image are indicated by their
source numbers from Table~\ref{tab:n4697x_src}
\label{fig:n4697x_raw_center}}
\end{figure*}

\begin{figure*}
\plotone{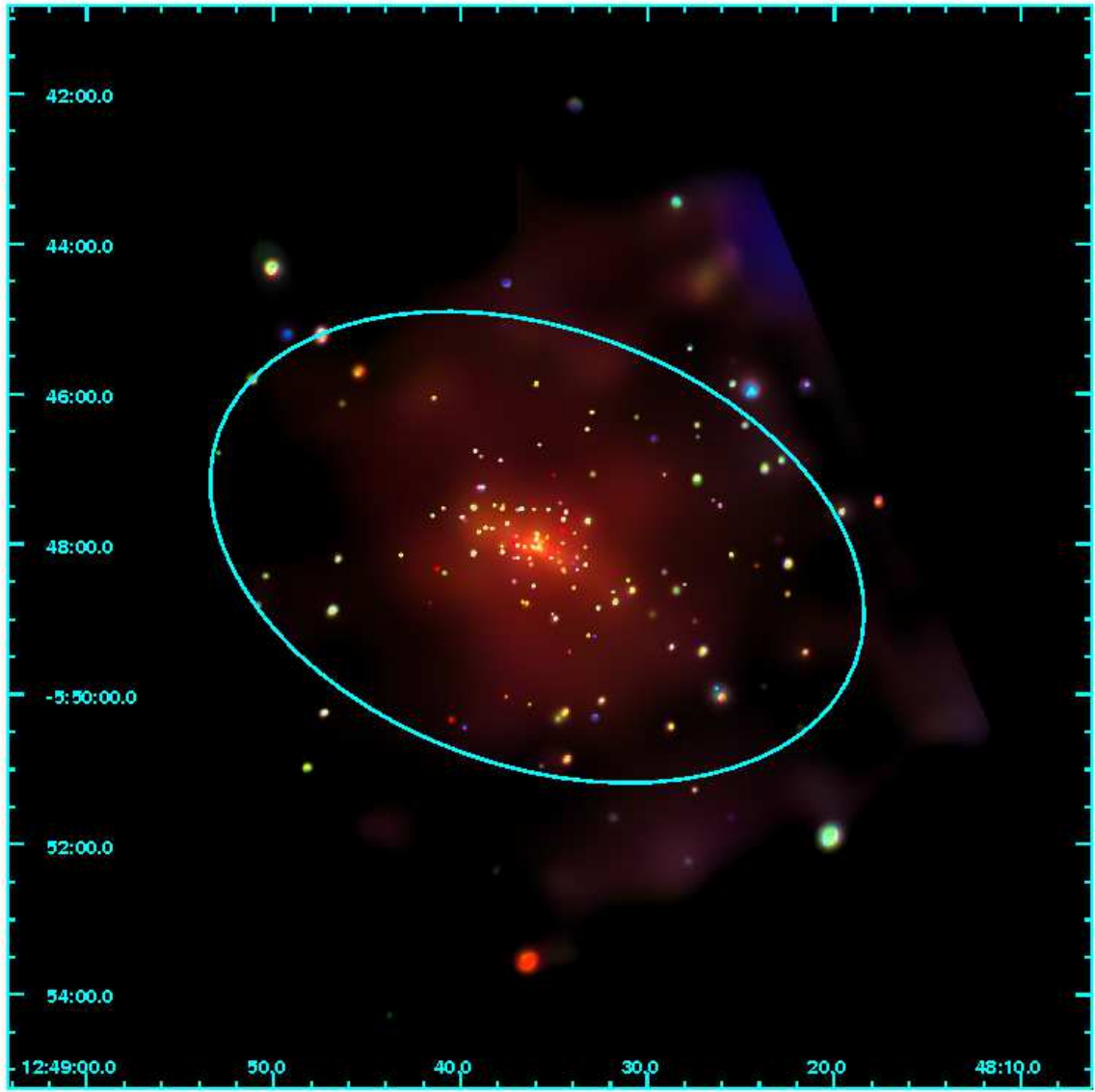}
\caption[Adaptively Smoothed Representative-Color X-ray Image of NGC~4697: S3 FOV]
{
Adaptively smoothed {\it Chandra} representative-color S3 image (with red =
0.3--$1 {\rm \, keV}$, green = 1--$2 {\rm \, keV}$, and blue = 2--$6 \, {\rm
keV}$) of NGC~4697 (all five observations), corrected for exposure and blank-sky
background. The intensity scale for the colors is logarithmic and ranges from $5
\times 10^{-7}$ to $1 \times 10^{-5} {\rm \, count \, s}^{-1}{\rm \,
arcsec}^{-2}$ in total surface brightness. The D25 ellipse is shown.
\label{fig:n4697x_adaptive_whole}}
\end{figure*}

\begin{figure*}
\plotone{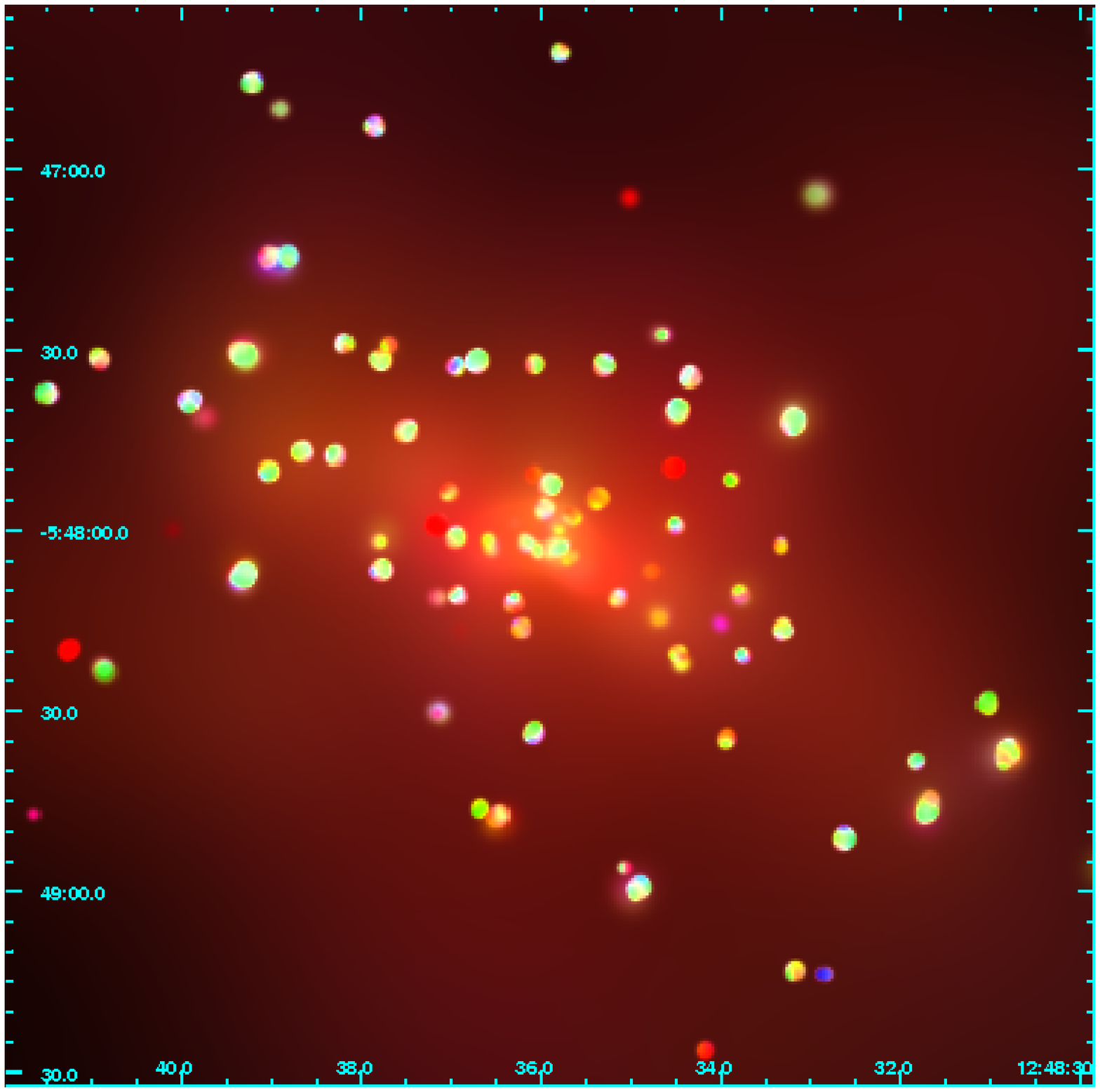}
\caption[Adaptively Smoothed Representative-Color X-ray Image of NGC~4697: Central $3\arcmin \times 3 \arcmin$ FOV]
{
Adaptively smoothed {\it Chandra} representative-color S3 image (with red =
0.3--$1 {\rm \, keV}$, green = 1--$2 {\rm \, keV}$, and blue = 2--$6
\, {\rm keV}$) of the central $3\arcmin \times 3 \arcmin$ of NGC~4697
(all five observations), corrected for exposure and blank-sky
background.
The color scaling is the same as in
Figure~\ref{fig:n4697x_adaptive_whole}.
\label{fig:n4697x_adaptive_center}}
\end{figure*}

In \citetalias{SIB2000} and \citetalias{SIB2001}, it was shown
that most of the X-ray emission in NGC~4697 is resolved into point
sources. We display the raw {\it Chandra} image from the combination
of all five observations in the 0.3--$6 {\rm \, keV}$ band in
Figures~\ref{fig:n4697x_raw_whole} and
\ref{fig:n4697x_raw_center}. These images are roughly consistent with
the previously published data; new sources have been detected due to
greater sensitivity, source variability, and increased FOV. The FOV of
the five observations, as well as a finding chart for the X-ray
sources, are overlaid on the raw images.

To contrast the detected sources with the diffuse emission, we
adaptively smoothed the {\it Chandra} S3 X-ray raw image using a
minimum signal-to-noise ratio (S/N) per smoothing beam of 3. We then
applied these smoothing scales to background-subtracted,
exposure-corrected images in the soft (0.3--$1 {\rm \, keV}$),
medium (1--$2 {\rm \, keV}$), and hard (2--$6 {\rm \, keV}$) bands.
The background includes the readout artifact in ACIS
and the deep blank-sky backgrounds compiled by
Maxim Markevitch\footnotemark[\ref{ftn:4697_bkg}].
We combined these three bands
to create a logarithmically scaled image between $5 \times 10^{-7}$ and
$1 \times 10^{-5} {\rm \, count \, s}^{-1}{\rm \, arcsec}^{-2}$
in total surface brightness.
The resulting representative-color images of the entire galaxy and of the central
$3\arcmin \times 3 \arcmin$ region are shown in
Figures~\ref{fig:n4697x_adaptive_whole} and
\ref{fig:n4697x_adaptive_center}.
There is clearly soft (red) diffuse emission near the center of NGC~4697.
The majority of the sources stand out clearly in color from the soft
diffuse gas, appearing yellow to green.
There are some soft sources which appear to
be associated with NGC~4697 based on their concentration towards the
center of the galaxy, while most of the hardest (blue) sources, which
are likely to be absorbed AGNs (see
\S~\ref{sec:n4697x_src_colors}), tend to lie in the outer regions of
the image.

\section{X-ray Source Detection}
\label{sec:n4697x_detections}

In Table~\ref{tab:n4697x_src}, we list all discrete sources detected
by {\sc wavdetect} over the 0.3--6~keV range. We ordered the sources
by increasing projected radial distance from the center of the galaxy,
$d$. Columns 1--8 provide the source number, IAU name, source position
(J2000), projected radial distance, projected semi-major distance from
the center of NGC~4697, $a$, photometric count rate with its
$1 \sigma$ error, and
S/N for the count rate.
Photon errors were calculated using the upper
Gehrels error approximation of $1+\sqrt{N+0.75}$ \citep{G1986}.
For comparison with \citetalias{SIB2001},
column 10 lists the source number used there. Notes for each source are
listed in column 11. The derived parameters and notes are expanded
upon in the text below.

To identify the discrete X-ray source population, we applied the
wavelet detection algorithm ({\sc ciao wavdetect} program) with
$\sqrt{2}$ scales ranging from 1 to 32 pixels with a source detection
threshold of $10^{-6}$.
Source detection was not done in regions with
an exposure of less then 10\% of the total for the observation.
We expect $\la 1$
false source (due to a statistical fluctuation in the background) for
each S3 image.
The source detections were first done on each observation separately to
create a
source list against which to register the astrometry.
We detected 97, 78, 87, 77, and 98 sources in Observations 0784, 4727,
4728, 4729, and 4730. We then registered the astrometry of each
Cycle-5 observation against Observation 0784. Using 55, 56, 51, and 55
sources matched to within $0\farcs5$, the relative astrometry
corrections were $0\farcs31$, $0\farcs38$, $0\farcs31$, and
$0\farcs48$.

To maximize S/N, we analyzed the wavelet detection results from the
combination of the five observations
(Figure~\ref{fig:n4697x_raw_whole}). Since there were five
approximately equally sensitive observations, we reduced the exposure
threshold of this merged source detection to 2\% of the total.
There was a 50\%
increase in the FOV compared to a single S3 chip, so we
expect $\la 1.5$ false source. We detected 158 sources. All of the
detections in Observation 0784 were in the merged detection list. In
the Cycle-5 observations, a few weak detections were not found in the
merged detection list. None of these detections would have fluxes
determined at the $\ge 3 \sigma$ level; detections not in the merged
detection list are not discussed further in this Paper.

Working from the coordinate list generated by {\sc wavdetect}, we used ACIS
Extract to create source extraction and masking regions, as well as refine the
source positions. For each observation, we created a source extraction region
consistent with the X-ray PSF at the source position. Most of the regions
encircled 90\% of the flux in the X-ray PSF at $\approx 1.5 {\rm \, keV}$. For
sources whose median photon energy over all five observations was not $\sim
0.6$--$2.6\, {\rm keV}$, we determined the PSF at either $\approx 0.3 {\rm \,
keV}$ (Sources 19, 23, 25, 62, and 78) or $\approx 4.5 {\rm \, keV}$ (Sources
96, 119, and 138). We used a lower percentage of the PSF in the case of few
sources whose regions would otherwise have overlapped in one of the observations
(85\% for Sources 2/5; 80\% for Sources 49/50; 50\% for Sources 72/73).
The {\sc wavdetect} source extents were compared to the PSF sizes at the
locations of the sources; all were consistent except those for Sources
8, 22, 123, and 158. Thus, these sources may be extended or multiple,
and are marked with a note in Table~\ref{tab:n4697x_src}. Masking
circular regions around the source at a radius encircling $>97\%$ of
the PSF were created for every source.

The refined source positions for a majority of the sources came from
the average position of 0.3--$6 {\rm \, keV}$ photons in the source
extraction regions. For Sources 142, 148, 152, and 154--158, whose
average positional offset from the optical axis is more than $5\arcmin$,
we correlated the 0.3--$6 {\rm \, keV}$ photons near the {\sc
wavdetect} coordinates against the average X-ray PSF of each source to
refine their positions.

To subtract out overlapping diffuse gas and background emission we
used a local background with an area approximately three times that of
each source's extraction region. The background region excluded
photons in the masking region. In cases where background regions
overlapped or fell along node/chip boundaries, we slightly altered
these overlapping regions, preserving the ratio of source to
background areas and ensuring that the source region and background
region had similar mean exposures.

The observed net count rates, their errors, and the S/Ns were calculated by
stacking the observations, correcting for background photons, and
dividing by the sum of the mean exposures over each source region.

We list the results of all analyses for all sources in this Paper's
tables; however, we restrict discussion of sources, except for
identification of possible optical counterparts, to the 126 that have
photometric count rates determined at the $\ge 3 \sigma$ level. These
significantly detected sources, hereafter the Analysis Sample, all
have at least 18 net counts.

The minimum detected count rate in the 0.3--$6 {\rm \, keV}$ band for
our Analysis Sample sources is $1.0 \times 10^{-4} {\rm \, counts \,
s}^{-1}$. This count rate is 2.6 times as deep as the count rate from
Observation 0784 alone \citepalias{SIB2001}. Three sources below the
Analysis limit (136, 152, 157) but with count rates above $10^{-4}
{\rm \, counts \, s}^{-1}$ are not covered by all five
observations. The other sources below the Analysis limit reach count
rates as low as $\sim 5 \times 10^{-5}{\rm \, counts \, s}^{-1}$.

We estimated the completeness of all sources through simulations
using {\sc marx 4.0.8}%
\footnote{See \url{http://space.mit.edu/CXC/MARX/}.}%
. We used the normalized background generated by {\sc wavdetect} and the
photometrically determined counts (after PSF corrections) to perform 400
simulated runs of our five observations of NGC~4697. The resulting completeness
correction factor, $c_{\rm sim}$, the ratio of the number of times a source was
simulated to the number of times it was detected, is accurate to $\sim 5\%$ for
the majority of sources; however, the uncertainty increases to $\sim 12\%$ for
the largest correction factors. All sources with observed count rates $>10^{-4}
{\rm \, counts \, ~s}^{-1}$ had correction factors $< 1.1$ except for Sources
72, 102, and 158. In the simulations, Sources 72 and 73 were often confused as
one source with a position closer to Source 73. Sources 102 and 158 had $c_{\rm
sim}=1.12$. The average completeness correction factor in the Analysis Sample is
$\sim 1.01$. For sources not in the Analysis Sample, $c_{\rm sim}$ averaged
$\sim 1.99$, and reached as high as 5.71 for the weakest sources. Our
completeness results are roughly consistent with \citet{KF2003}.

There are objects unrelated to NGC~4697 among the detected sources. Since the
FOV sampled by the ChaMP survey \citep{KWG+2004} is larger than the {\it
Chandra} deep fields, the source counts from the former should be less
susceptible to cosmic variance. Therefore, we chose to use their soft band
source counts to estimate the number of foreground or background objects at
different flux levels. At the flux limit of our Analysis Sample, $\sim 9 \times
10^{-16} {\rm \, erg \, cm}^{-2} {\rm \, ~s}^{-1}$, we expect $\approx 29$
foreground or background objects, including corrections for exposure and
completeness. We estimate $\approx 7$ of the sources outside of the Analysis
Sample are also unrelated to NGC~4697. Since these sources should be fairly
uniform over the FOV, sources close to NGC~4697 are more likely to be associated
with the galaxy than sources farther out.

\section{Optical Counterpart Identification}
\label{sec:n4697x_opt_ids}

\subsection{Existing Catalog Identifications}
\label{sec:n4697x_cat_ids}

The refined source positions of seven X-ray sources from the combined S3 image
agreed with positions from the Tycho-2 Catalog
\citep{HFM+2000},
the 2MASS Point Source and Extended Source Catalogs \citep{SCS+2006},
and/or
the USNO-B Catalog \citep{MLC+2003}.
(When a source appeared in both 2MASS catalogs, the Point Source Catalog
positions were used.)
In cases where there are counterparts in multiple catalogs,
we adopt the positions in the order of the catalogs listed above.
These seven sources (15, 55, 84, 117, 149, 155, and 156) were used to
check the absolute astrometry.
Since the mean positional offset of the
sources was $0\farcs07\pm0\farcs30$ in R.A. and
$0\farcs10\pm0\farcs27$ in Dec., the ACIS Extract positions are
consistent with no required absolute astrometric change. The typical
absolute astrometric errors are probably $\sim 0\farcs4$ near the
field center, with larger errors for weaker sources with extended
PSFs.

Having established the absolute astrometry, we conservatively
considered all optical sources within $2\arcsec$ as potential optical
counterparts. We summarize their optical properties in
Table~\ref{tab:n4697xo_gb_src}. The first three columns list the X-ray
source number, designation of the optical counterpart, and positional
offset between the X-ray and optical catalogs. In the fourth column,
we list the photometric properties of the optical counterpart, while
we list notes about the optical properties in the fifth column. We
classify the counterpart as optically extended or an optical point
source. For 2MASS objects, we use its values of the reduced $\chi^2$
for fitting PSFs to each source in each
band (a reduced $\chi^2 > 2$
indicates the optical counterpart is extended, and may be a
galaxy). For the USNO-B1 objects, we use their star-galaxy separation
class (objects in the lower half of classes are classified as
optically extended). We add a question-mark when we are unsure of the
classification. Typically this occurs because
galaxy light may have contaminated the analysis or because the
classification in at least one color differs significantly from the
other colors.

In addition to the sources used to check astrometry, Sources 1, 8, and
118 have potential optical counterparts. Since Source 1 is $0\farcs5$
away from the adopted center of NGC~4697, it may be a central AGN.
Since this source could also be an LMXB (or several confused LMXBs)
near the center of the galaxy, we did not use this match to check the
astrometry. The optical counterparts of Sources 8, 15, 117, 118, and
155 appear to be optically extended. We note that Source 117 is a
known AGN \citepalias{SIB2001}. Although Source 118 is associated with
\object{2MASX J12483504-0550473} in NED, that source is actually
$12\arcsec$ away and its counterpart, USNO-B1 0841-0238567, is clearly
different on Digital Sky Survey (DSS) images. Since the extrapolated
fiducial radius of \object{2MASX J12483504-0550473} is $6\farcs2$, it
is unlikely that Source 118 is associated with that galaxy. Sources
55, 84, 149, and 156 appear point-like from the ground; Source 156 is
clearly the bright foreground star \object{BD-05 3573}. Finally, we
note that Sources 103, 123, 132, 143 may have uncatalogued
counterparts on DSS second generation images. Footnotes in
Table~\ref{tab:n4697x_src} indicate X-ray sources with possible
optical counterparts.

In \citet{SIB2001}, associations with GCs in lists of \citet{H1977} and
Kavelaars (2000, private communication) were made. Sources 101 and 117 were
previously identified with \citet{H1977} GC candidates; we find no new GC-LMXB
candidates among the \citet{H1977} sources and note that Source 117 and its
optical counterpart has already been shown to actually be an AGN. Sources 79,
80, 83, 84, 86, 88, 93, 98, 101, 109, 113, and 114 are associated with GC
candidates in the Kavelaars data, with Sources 79, 86, 88, 93, and 113
representing new detections. Some of the Kavelaars GC candidates with potential
X-ray counterparts are also in the HST-ACS FOV discussed in
\S~\ref{sec:n4697x_hst_ids} (Sources 79, 80, 83, 84, 86). The measured colors
and sizes from the HST-ACS observations allow us to more accurately identify GC
candidates. The colors of Source 84 suggest it is not a GC, while the other
sources have colors and sizes consistent with GCs. Since the Hanes and Kavelaars
GC datasets do not have the same sensitivity or measured properties of the
HST-ACS dataset or of our recently obtained flanking field ACS observations, we
will not make further use of GC counterparts outside of the HST-ACS central FOV
here.

We matched the X-ray sources with a previous {\it ROSAT-HRI} X-ray
observation of NGC~4697 \citep[][hereafter
\citetalias{ISB2000}]{ISB2000}. Twelve sources within $4\arcmin$
of NGC~4697 were detected by {\it ROSAT-HRI} with $L_X \gtrsim 2.6
\times 10^{38} {\rm \, erg \, s}^{-1}$, after correcting for the
distance and spectra we use in this Paper. We match
\citetalias{ISB2000} Sources 1, 2, 4, 5, 6, 9, 10, 11, and 12 with
this Paper's Sources 103, 80, 69, 54, 118, 58, 128, 117,
143. Confusion in the center of NGC~4697 limits matching
\citetalias{ISB2000} Source 7 with this Paper's Sources 1, 3, 4, and
6; although it is likely matched to this Paper's Source 1, which is
the brightest of the four sources. Similarly, confusion limits
matching \citetalias{ISB2000} Source 8 with this Paper's Sources 6, 7,
10, and 11. We clearly do not detect
\citetalias{ISB2000} Source 3.

We also matched sources with \citetalias{SIB2001}, whose data
(Observation 0784) is a subset of the data in this Paper. We list
those matches in Table~\ref{tab:n4697x_src}. Only two
\citetalias{SIB2001} sources (7 and 87) are undetected in the larger
dataset. \citetalias{SIB2001} Sources 7 and 10 are separated by only
$1\farcs1$. It is unclear whether they are a single source.
\citetalias{SIB2001} Source 87 was near the limit of the S/N in
\citetalias{SIB2001}. It was not detected in the individual
reanalysis of Observation 0784, even when detections were made in the
0.3--10~keV range to match \citetalias{SIB2001}.

\subsection{HST-ACS Identifications}
\label{sec:n4697x_hst_ids}

In our HST-ACS observation, we placed the center of NGC~4697
$\sim 20\arcsec$ from the center of the ACS FOV to avoid a chip
boundary. Within this FOV, we can identify optical sources and
separate out the GCs using a combination of magnitude, color, and
spatial extent. This method is discussed in detail in \citet{PJC+2006}
and Jord\'{a}n et al.\ (2008, in preparation). Out of 703 optical
sources, there are 298 GC candidates.

We registered the HST observation to the Chandra observation using all 703
optical sources. Through cross-correlation techniques, we determined that the
HST coordinates required astrometric shifts of $-0\farcs27$ in R.A. and
$-0\farcs42$ in Dec. Based on singly matched sources within $1\arcsec$, we
estimate a relative astrometric error of $0\farcs21$ and $0\farcs25$ in R.A. and
Dec., respectively, or $0\farcs32$ in quadrature. Since most of these matches
occur close to the X-ray pointing center, it is not surprising that this
astrometric error is slightly smaller than the astrometric error derived from
ground-based catalogs. After correcting the astrometry, we determined that the
ACS FOV covers X-ray Sources 1--67, 69--87, 89, 94, and 108. Due to a prominent
elliptical dust feature ($\approx 7.0 \arcsec \times 1.7 \arcsec$ or $\approx
384 {\rm \, pc} \times 93 {\rm \, pc}$), we did not attempt to detect optical
sources in the region of X-ray Sources 1--5 and 8. X-ray Source 68 falls in the
chip boundary. Adopting a search radius of $1\arcsec$, we find 42 optical
counterparts to 39 X-ray sources; three X-ray sources have two candidate optical
counterparts within $1\arcsec$. Thirty-three of the X-ray sources are associated
with a single GC counterpart, while one X-ray source is likely to be associated
with one of two potential GC counterparts. We list the optical properties of the
matched sources in Table~\ref{tab:n4697xo_hst_src}.

By randomizing the P.A. of a source list, assuming both a circular profile and a
profile matching the galaxy's elliptical isophotes, we have determined the
percentage of false matches within a given radius. Randomizing the optical
source list, we find that $\approx 6.8\%$ of X-ray sources without a physically
associated optical source will have a false optical match within $1\arcsec$. 
This is consistent with only 3/39 of the X-ray/optical matches being false. If
one considers only the GCs, $\approx 4.2\%$ of X-ray sources will have a false
match. This is consistent with only 2/34 of the X-ray/GC matches being false. 
Randomizing the X-ray source list, we find that $\approx 0.73\%$ of GC sources
without a physically associated X-ray sources will have a false optical match
within $1\arcsec$. Finally, randomizing the optical source list and comparing to
the unrandomized optical positions suggests that $\approx 1.8$ X-ray sources
will be matched to two optical sources within $1\arcsec$ by chance; we find
three such matches.

\section{GC/LMXB Connection}
\label{sec:n4697x_gclmxb}

\begin{figure}
\plotone{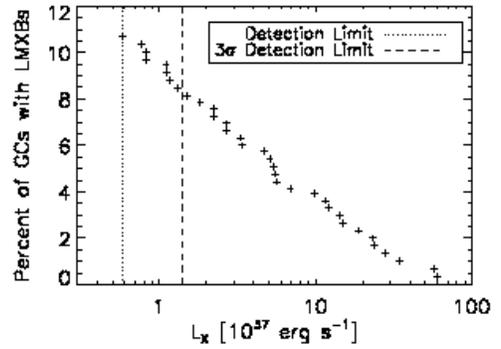}
\caption[Percentage of GCs with an LMXB vs.\ Limiting X-ray Luminosity in NGC~4697]{
Percentage of GCs with an LMXB within $1\arcsec$, corrected for random
associations, as a function of the limiting X-ray luminosity.
The detection limits for all sources and for the Analysis
Sample ($3 \sigma$) are indicated.
\label{fig:n4697x_frac_gc_w_lmxb}}
\end{figure}

Having identified the LMXBs associated with GCs and determined the percentage of
falsely matched sources, we now explore the GC/LMXB connection. The broadest
measures of the GC/LMXB connection are the fraction of LMXBs associated with GCs
and the fraction of GCs associated with LMXBs. The fraction of LMXBs associated
with GCs, $f_{X,{\rm GC}}$, is $38.4^{+6.1}_{-5.7}\%$ and does not appear to
depend on X-ray luminosity. On the other hand, the fraction of GCs with an LMXB,
$P_{X}$, naturally depends on the limiting X-ray luminosity. The result is shown
in Figure~\ref{fig:n4697x_frac_gc_w_lmxb}. In NGC~4697, and other early-type
galaxies \citep{SKI+2003}, $P_{X} \approx 4\%$ above $10^{38} {\rm \, ergs \,
s}^{-1}$. At the limit of our Analysis Sample, $1.4 \times 10^{37} {\rm \, ergs
\, s}^{-1}$, $P_{X}$ has increased to $8.1^{+1.9}_{-1.6}\%$ ($\sim 2.7 \times
10^{-7}$ LMXBs per $L_{\odot,z}$ normalizing LMXB detection to GC luminosity in
the $z$ band). Among all detected sources ($L_{X} > 0.6 \times 10^{37} {\rm \,
ergs \, s}^{-1}$), $P_{X}$ rises to $10.7^{+2.1}_{-1.8}\%$ ($\sim 3.5 \times
10^{-7}$ LMXBs per $L_{\odot,z}$). Since we are incomplete below the limit of
our Analysis Sample and active LMXBs have $L_X \gtrsim 10^{36} {\rm \, ergs \,
s}^{-1}$, it is likely that the percentage of GCs with an active LMXB is even
higher.

\begin{figure*}
\plottwo{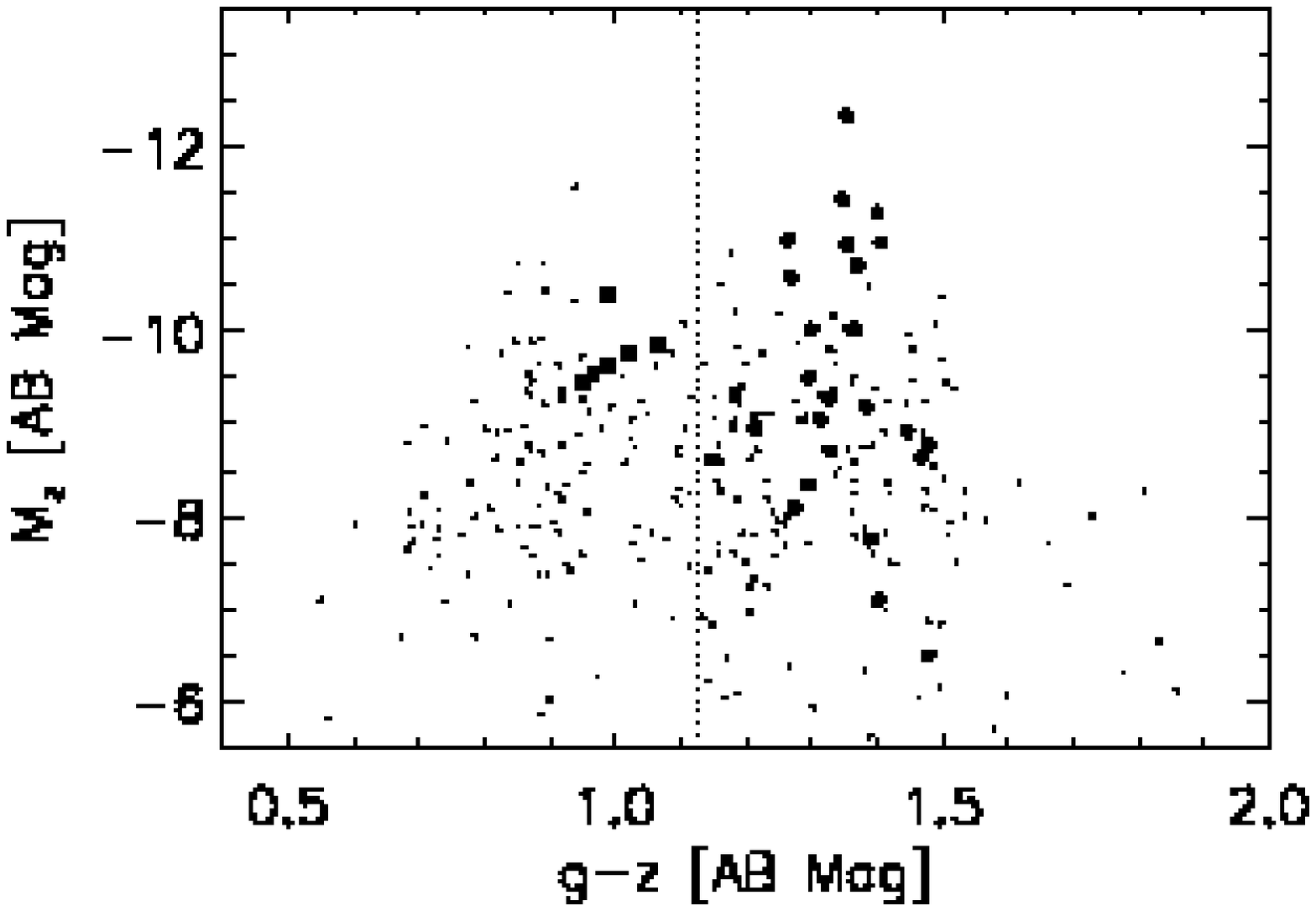}{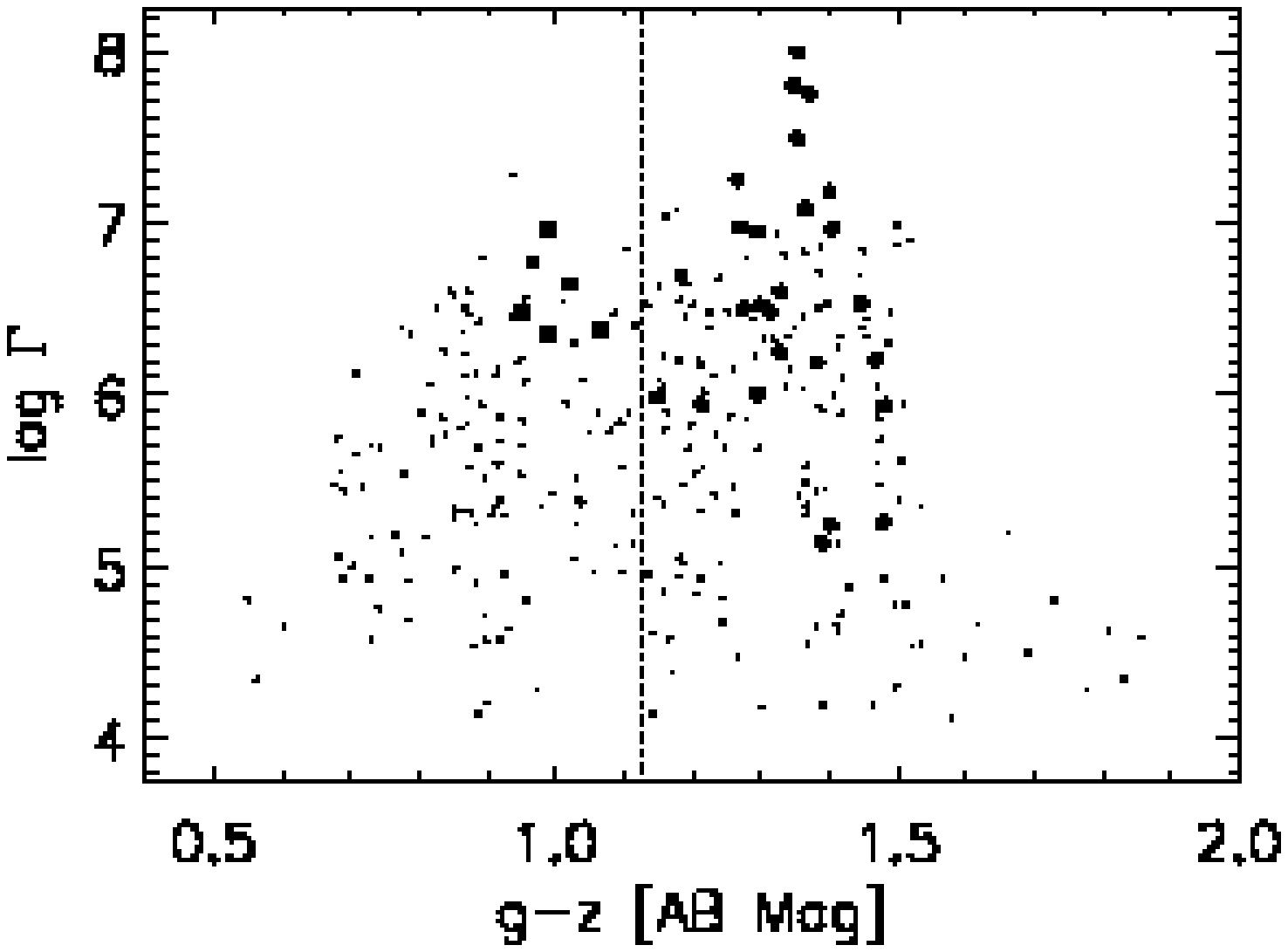}
\caption[Color Magnitude Diagram of GCs in NGC~4697]
{
Color ($g-z$) - magnitude ($M_{Z}$) diagram ({\it left})
and color - encounter rate ($\Gamma_{h} \propto M^{3/2} \, r_{h}^{-5/2}$)
diagram
({\it right})
for GCs in HST-ACS.
The larger symbols indicate GCs containing
LMXBs.
The vertical line indicates the separation of GCs into
blue and red GC populations following \citet{PJC+2006}. LMXBs reside
more often in GCs with larger optically luminosities, larger encounter
rates, and redder colors.
\label{fig:n4697x_gc_cmd}}
\end{figure*}

The optical properties of GCs are known to affect the GC/LMXB connection; LMXBs
are found more often in optically bright GCs and in red GCs \citetext{e.g.,
\citealt{KMZ+2003}; \citealt{SKI+2003}; \citealt{JCF+2004}; \citealt{KMZ2007};
\citealt{SJS+2007}, hereafter \citetalias{SJS+2007}}. The GC/LMXB connection in
the center of NGC~4697 clearly follows this now familiar pattern
(Figure~\ref{fig:n4697x_gc_cmd}, {\it left}). The color dependence is often
interpreted as a metallicity dependence \citetext{\citetalias{SJS+2007};
\citealt{KMZ2007}}. In addition, the size of a globular cluster also plays a
role, with smaller GCs being more likely to host LMXBs
\citetext{\citetalias{SJS+2007}; \citealt{JSM+2007}}. Following \citet{SKI+2003}
and \citetalias{SJS+2007}, we have compared the distributions of GCs with and
without LMXBs using the two-sample Kolmogorov-Smirnov test ($P_{\rm KS}$), which
measures the probability that two populations sampled from the same distribution
have a maximum difference at least as large as is observed, and the
non-parametric Wilcoxon rank-sum test \citep[$\sigma_{\rm WRS}$, equivalent to
the Mann-Whitney rank-sum test;][]{MW1947}, which measures the probability that
a random sampling of two distributions with the same median would produce at
least the observed difference in the sum of the ranks of the two distributions.

First, we compared the optical luminosities of GCs (represented by their $M_z$)
with and without LMXBs. The mean optical magnitude of GCs with LMXBs is $1.1
{\rm \, mag}$ brighter then GCs without LMXBs, and we found $P_{\rm KS} = 2.6
\times 10^{-4}$ and $\sigma_{\rm WRS}=4.5$. That is, LMXBs are preferentially
associated with optically luminous GCs at high statistical significance. Next,
we compared the $g-z$ color distributions of GCs with and without LMXBs. The
mean color of GCs containing LMXBs is nearly 0.2 magnitudes redder and the
distributions are clearly different ($P_{\rm KS} = 3.6 \times 10^{-4}$ and
$\sigma_{\rm WRS}=3.0$). These results are illustrated in
Figure~\ref{fig:n4697x_gc_cmd} ({\it left}). Quantifying the probabilities that
GC contain LMXBs, we find LMXBs in $4.8^{+2.7}_{-1.9}\%$ of the blue GCs ($\sim
2.1 \times 10^{-7}$ LMXBs per $L_{\odot,Z}$) and $15.1^{+3.2}_{-2.8}\%$ of the
red GCs ($\sim 4.9 \times 10^{-7}$ LMXBs per $L_{\odot,Z}$). The relative ratio
of the specific frequencies for GC-LMXBs in red-GCs versus blue-GCs is smaller
than the ratio of the probabilities a red-GC versus a blue-GC contains an LMXB. 
Although KS tests and Wilcoxon rank-sum test indicate the optical luminosities
of red and blue GCs are likely to be drawn from the same parent distribution,
the mean luminosity of the red-GCs is larger than the blue-GCs. We note that the
mean luminosities of the GCs are strongly affected by the numbers of the most
luminous GCs.

\citet{JCF+2004} used HST-ACS data from M87 to compare the GC/LMXB
connection with the encounter rates due to tidal capture and exchange
interactions. The encounter rates are proportional to $\Gamma \equiv
\rho_0^{1.5} \, r_{\rm c}^2$, where $\rho_0$ is the central density, and $r_{\rm
c}$ is the core radius. These values were estimated from fitting King profiles
to the GC light and applying the Virial Theorem to relate the velocity of stars
to the core radius. They found a strong indication that dynamical processes play
a key role; however, the somewhat large uncertainties in the concentrations (the
ratio of the tidal radius to the core radius, which is uniquely related to the
ratio of the half-light radius to the core radius for King models) make
estimates of $\Gamma$ less attractive observationally as a measure of encounter
rates. Since the half-light radii, $r_{h}$, and surface brightnesses of GCs,
$\sigma_{h} \propto M \, ( r_{h})^{-2}$, are better constrained, we choose to
use an alternate measure of the encounter rate, $\Gamma_{h} \equiv
\sigma_{h}^{3/2} r_{h}^{1/2} \propto M^{3/2} r_{h}^{-5/2}$. In recent
observations of Centaurus A, a much closer elliptical galaxy, \citet{JSM+2007}
showed that concentration does not seem to be a fundamental variable in
determining the presence of LMXBs in GCs, with the more fundamental parameters
being related to central density and size. While this implies that the LMXB-GC
connection should be stronger for $\Gamma$ than $\Gamma_h$, results for
$\Gamma_h$ can be taken as representative of the more fundamental parameters of
central density and size. Mass segregation of GCs with the same half-mass radius
\citep{J2004} has been suggested as a possible explanation for a correlation
typically observed between half-light radius and color
\citep[e.g.,][]{KW2001,JCB+2005}, with redder GCs being smaller. However, that
correlation may also represent the convolution of an underlying
size-galactocentric distance relation and the different spatial distribution of
the metal-poor and metal-rich subpopulations of GCs \citep{LB2003,SLS+2006};
this is the so-called ``projection effect''. Given that the observed correlation
of half-light radii with color might not reflect a corresponding correlation for
half-mass radii, we consider an alternative to the directly measured half-light
radii that corrects for the empirical size-color correlation; we call this
corrected radii ``half-mass'' radii, $r_{h}^{\prime}$. We follow equations 1, 2,
5 , and 12 of \citetalias{SJS+2007} to calculate the mass, half-mass radius,
encounter rate proxy $\Gamma_{h}$, and metallicity $Z$ of our GCs. We note here,
and throughout this section, that the choice of which radius to use ($r_h$ or
$r_{h}^{\prime}$) mainly alters results for the metallicity, as opposed to the
size.

In Figure~\ref{fig:n4697x_gc_cmd} ({\it right}), we display the color -
encounter rate diagram. Globular clusters with larger encounter rates are more
likely to contain LMXBs; the hypothesis that the distributions of encounter
rates are the same for GCs with and without LMXBs is rejected ($P_{\rm KS} = 1.7
\times 10^{-7}$ and $\sigma_{\rm WRS}=5.8$). This picture does not qualitatively
change if we calculate the encounter rates with the half-mass radii ($P_{\rm KS}
= 5.8 \times 10^{-7}$ and $\sigma_{\rm WRS}=5.5$). If LMXBs are preferentially
found in core-collapsed GCs, the difference between the actual encounter rates
of GCs with and without LMXBs will be larger than what we measured. We note that
the concentrations of GCs with and without LMXBs do not differ when considering
samples with matched masses \citep{JSM+2007}, statistically validating the use
of $\Gamma_h$ as a proxy for encouter rate. The correlation with encounter rate
is actually stronger than that found for the optical luminosity, because smaller
GCs appear (marginally) more likely to contain GCs in this galaxy, consistent
with Centaurus A \citep{JSM+2007} and the Virgo elliptical galaxies as a whole
\citep{SJS+2007}. When we compare the distributions of $r_h$ for GCs with and
without LMXBs, the hypothesis that they are the same is rejected ($P_{\rm KS} =
7.3 \times 10^{-3}$ and $\sigma_{\rm WRS}=3.6$); however, part of this
correlation is due to the fact that redder GCs are both more likely to contain
LMXBs due to their color, and are also smaller for the same mass. The
differences between GCs with and without LMXBs are rejected more marginally when
comparing the distributions of $r_h^{\prime}$ ($P_{\rm KS} = 2.2 \times 10^{-2}$
and $\sigma_{\rm WRS}=2.6$).

Recently, \citet{KKF+2006} suggested that $P_{X}$ may have a spatial dependence
due to higher encounter rates in GCs near the centers of their host galaxies. 
Although our comparison of the projected galactocentric distances $d$ of GCs
with and without LMXBs does not indicate that they are drawn from two separate
populations at a significant level ($P_{\rm KS} = 0.21$ and $\sigma_{\rm
WRS}=1.6$), we do note that the median galactocentric distance of GCs with LMXBs
is smaller than those without. This question will be best addressed with our
HST-ACS observations of the entire galaxy.

As in \citetalias{SJS+2007}, we have attempted to fit the expected number
($\lambda$) of LMXBs in a GC with the sum of the false number of $\lambda_f$
LMXBs matched to a GC and the true expected number $\lambda_t$ parameterized by
power-law dependences on the GC properties. We considered the forms:
\begin{equation}
\lambda = \lambda_f +
          A \, M^{\alpha} \,
               (Z/Z_\odot)^{\beta} \,
               (r)^{\delta} \, ,
\label{eq:expected1}
\end{equation}
and
\begin{equation}
\lambda = \lambda_f +
          A \, (Z/Z_\odot)^{\beta} \,
               \Gamma^{\epsilon},
\label{eq:expected2}
\end{equation}
where we evaluate the parameters separately using both $r_h$ and $r_h^\prime$.
The expected number of LMXBs in a GC can be converted to a probability that
there are no LMXBs, $P_{nX,i} = e^{-\lambda_i}$, and the probability that there
is at least one LMXB, $P_{X,i} = 1- e^{-\lambda_i}$. One can then maximize the
log likelihood for a given form of $\lambda$, $\psi = \ln [(\underset{nX}{\prod}
P_{nX}) \ (\underset{X}{\prod} P_{X})]$, where the products are taken over the
lists of GCs with no LMXBs and GCs with LMXBs (Table~\ref{tab:n4697gclmxb_par}).

Although the general maximum likelihood statistic does not provide a measure of
goodness-of-fit, relative improvements ($\Delta \psi = -\Delta \chi^2 /2$) can
be used to determine whether a given fit is a statistical improvement over a
previous fit given the change in the number of degrees of freedom (dof). The
change in the log-likelihood can also be used to provide errors on the fitted
power-law indices. Here, we assumed that the errors in the fitting parameters in
equations~(\ref{eq:expected1}) \& (\ref{eq:expected2}) are much larger than the
uncertainties in the GC parameters ($M$, $Z$, and $r_{h}^\prime$) due to either
measurement errors or systematic errors in the conversions. Given the derived
sizes of our errors, this is justified. To use the relative change in the
log-likelihood ($\Delta \psi$), we first established a baseline value of the
log-likelihood ($\psi_0$) for the case where the expected number of LMXBs
$\lambda$ was constant and did not depend on any GC properties
(Table~\ref{tab:n4697gclmxb_par}, row 1).

We then fit various combinations of dependencies of $\lambda$ on GC properties,
and determined the values of $\psi$ for the best-fit values. In rows 2--5 of
Table~\ref{tab:n4697gclmxb_par}, the expected number of GCs $\lambda$ is assumed
to depend on only a single GC property (the mass, metallicity, half-light
radius, or half-mass radius). Each fit is significantly better ($\Delta \chi^2 =
\, $$-$27.5, $-$10.4, and $-$7.8 for one less dof) than the baseline fit,
suggesting that all three properties affect $\lambda$, with mass having the
strongest effect.

In the next two rows (6 and 7), the expected number of LMXBs is assumed to
depend only on the encounter rate parameter $\Gamma_h$. This fit was
considerably better than those for any other single parameter. Thus, it appears
that the most important single factor determining the occurrence of LMXBs in GCs
is the dynamical encounter rate. Given that the encounter rate calculated from
the half-light radius includes some metallicity dependence, it is not surprising
that it fits better than the encounter rate calculated from the half-mass
radius.

In the next 7 rows (8--14), the expected number of LMXBs is assumed to depend on
pairs of the GC properties. It is particularly interesting to compare rows 8 and
9 with rows 6 and 7. The encounter rate parameter $\Gamma_h$ is calculated from
the mass $M$ and the half-light (half-mass) radius $ r_h (r_h^\prime)$ (eq. 5 of
\citetalias{SJS+2007}). Thus, these two sets of rows compare a general
dependence on mass and radius with the specific form expected if LMXBs are
formed dynamically in GCs. The separate mass and half-light radius dependence
produces a marginally statistically significant better fit ($\Delta \chi^2 =
3.0$, which implies $92\%$ significance), while the separate mass and half-mass
radius dependences produces a less significant fit ($\Delta \chi^2 = 1.8$, which
implies $82\%$ significance). However, both are significantly better fits than
just the GC mass, metallicity, or size dependence alone. In rows 8 and 9, note
that the best-fit exponents for the mass ($\alpha = 1.28 {\rm \ and \ } 1.31$)
are very close to the value predicted by the dependence on $\Gamma_h$ in rows 6
and 7 ($\alpha = 1.5 \times \epsilon = 1.34$ and $1.31$), although the
dependences on radius are steeper ($\delta = -3.35$ and $-3.19$) than predicted
by rows 6 and 7 ($\delta = -2.5 \times \epsilon = -2.22$ and $-2.18$).

In rows 10--12, the metallicity dependence and either mass or radius dependence
are allowed to vary. Varying mass and metallicity dependences have the strongest
effect; however, neither effect is as strong as varying the dynamical
dependences (combination of mass and radius).

When we compare fits including metallicity and sizes calculated by half-light
radii to metallicity and sizes calculated by half-mass radii (rows 11 and 12, 13
and 14, 15 and 16, and 17 and 18), we see a consistent pattern where the index
involving size barely changes, but the metallicity index is smaller when
half-light radii are used. Although this difference is not large for most of
these comparisons compared to their precision, the accuracy of the metallicity
index depends critically on which size is used. Under the mass-segregation
hypothesis for the GC color-size dependence \citep{J2004}, the half-mass radius
is clearly the correct size to use. We argue that this choice is also correct
under the projection effect hypothesis \citep{LB2003}; however in this case, the
correction to ``half-mass'' is actually a correction that is removing an effect
on galactocentric distance. We note that the current analysis for NGC~4697 is
insufficient to test this hypothesis.

In rows 13--18, we combine dynamical dependence and metallicity dependence
variations, providing the best fits to the data. We adopt row 14, dependence on
metallicity and encounter rate calculated by half-mass radius, as our best fit:
$\lambda_t \propto \Gamma_{h}^{0.79^{+0.18}_{-0.15}} \,
(Z/Z_\odot)^{0.50^{+0.20}_{-0.18}}$. In addition to being the best statistical
fit, we also note that this fit uses our preferred half-mass radius in
calculating the encounter rate. The dependence on interaction rate matches well
with the Galactic value found by \citet{PLA+2003} of $\lambda \propto
\Gamma^{0.74\pm0.36}$. As NGC~4697 is a significant subset of the data
used in \citetalias{SJS+2007}%
\footnote{We note that there was a minor error in the $r_h$ of NGC~4697 used in
\citetalias{SJS+2007}, but that the changes this causes are within the
quoted errors.},%
it is unsurprising that it matches well to the $\lambda \propto
\Gamma_h^{0.82\pm0.05}$ relation determined for the GC/LMXB connection
in Virgo elliptical galaxies. For comparison with \citet{JCF+2004}, we also fit
the form $\lambda_t \propto \rho_0^{1.35^{+0.26}_{-0.21}} \,r_{\rm c}^2 \
(Z/Z_\odot)^{0.50^{+0.20}_{-0.19}}$ (row 16); our results agree. We note that
allowing mass, metallicity, and size to vary all at once (rows 17 and 18) does
not significantly improve the fits compared to allowing encounter rates and
metallicity to vary (rows 13 and 14).

For our best fit, we find
\begin{equation}
\label{eq:n4697x_lambda}
\lambda_t = 3.0 \times 10^{-6} \,
           \Gamma_{h}^{0.79^{+0.18}_{-0.15}} \,
           (Z/Z_\odot)^{0.50^{+0.20}_{-0.18}},
\end{equation}
where $\Gamma_h$ is calculated from $r_h^\prime$. Since our analysis includes
matches below the completeness limit, the measured normalization is intermediate
between the true normalizations at the detection limit and the completeness
limit. We adopted the best-fit parameters of equation \ref{eq:n4697x_lambda} and
only included matches above the completeness limit to determine that the
normalization at the completeness limit of $1.4 \times 10^{37} {\rm \, ergs \,
s}^{-1}$ is $2.1 \times 10^{-6}$. We can use the normalizations to estimate the
number of GCs that might contain multiple LMXBs.
\citep[e.g., there are two Galactic LMXBs in M15;][]{WA2001}.
By summing $1 - \exp^{-\lambda} ( 1 + \lambda )$ over all the GCs, we calculate
that $\sim7$ and 4 GCs contain multiple LMXBs above the detection and
completeness limits, respectively. The majority of the 34 GCs we detect with
LMXBs are likely to contain only one LMXB.

\section{X-ray Luminosities and Luminosity Functions}
\label{sec:n4697x_src_lum}

We used the best-fit Chandra X-ray spectrum of the inner resolved
sources ($a < a_{\rm eff}$: Table~\ref{tab:spectra_n4697}, row 3 below) and the
assumption that each source was at the distance of
NGC~4697 to convert
the observed source count rates into unabsorbed X-ray (0.3--10 keV)
luminosities ($L_X$).
The fluxes were corrected for exposure (including vignetting), the time
dependent QE degradation of the ACIS-S3 chip, and the PSF fraction of source
counts within the region used to extract the counts. For a typical source, the
individual conversion factors from observed count rates were 1.18, 1.50, 1.46,
1.46, and $1.47 \times 10^{41} {\rm \, ergs
\, count}^{-1}$, for each of the observations ordered by time.
We list the individual luminosities
of Observations 0784 (A), 4727 (B), 4728 (C), 4729 (D), and 4730 (E)
in columns 2--6 of Table~\ref{tab:n4697x_luminst}.

\begin{figure}
\plotone{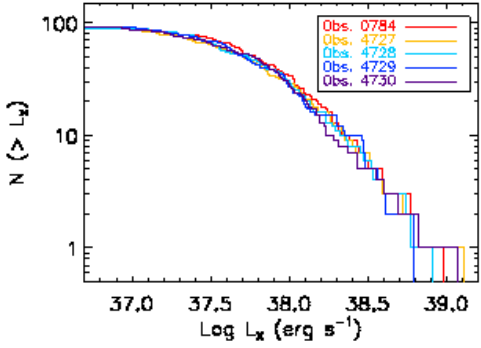}
\caption[Cumulative Luminosity Functions of Discrete X-ray Sources in NGC 4697: Individual Observations]{
Cumulative luminosity functions for each of the five observations of
the sources detected at the $> 3 \sigma$ level within $a <
220\arcsec$.
The two-sample K-S tests do not indicate that any two of the observations
are drawn from different populations.
\label{fig:n4697x_lf_ind}}
\end{figure}

In Figure~\ref{fig:n4697x_lf_ind}, we
display the individual observation LFs of the sources in the Analysis Sample
within $a < 220\arcsec$.
We calculate the probability the LFs of each pair of observations are
drawn from the same population using the two-sample K-S test. Since
$P_{\rm KS}$ range from 0.33 (0784 versus 4727) to
0.94 (4729 versus 4730), we believe that the LFs do not change
significantly on our inter-observation timescales, which vary from
$11 {\rm \, d}$ to $4.6 {\rm \, yr}$.

We combined the luminosities from all of the different observations;
as before, we included the effects of varying exposure (including vignetting),
the time dependent QE degradation of the ACIS-S3 chip, and the PSF fraction of
source counts within the region used to extract the counts.
We give this luminosity, combining all five
observations, as $L_{\rm all}$ in Table~\ref{tab:n4697x_lumcolor}
(column 4).

\begin{figure}
\plotone{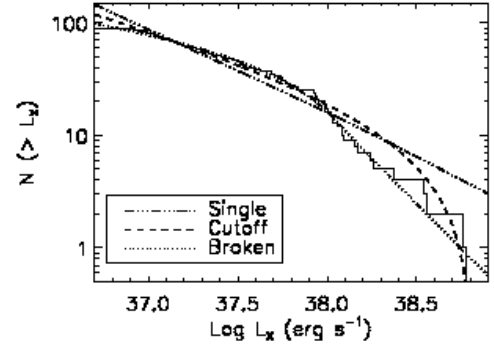}
\caption[Cumulative Luminosity Functions of Discrete X-ray Sources in NGC 4697: Constant Sources]
{
Cumulative luminosity function of the constant sources
($L_{X}=L_{\rm constant}$) detected within $a < 220\arcsec$ using
their cumulative luminosity for all observations.
The continuous curves are the sum of the
best-fit LMXB luminosity functions (to sources detected at the
$3 \sigma$ level) and the expected background source counts. The broken
power law is the best fit model; however, a cutoff power law is acceptable
according to the K-S test.
\label{fig:n4697x_lf_constant}}
\end{figure}

One can compare $L_{\rm all}$ to the individual luminosities
using $\chi^2$ to test if the individual values are all consistent with a
constant luminosity.
In Figure~\ref{fig:n4697x_lf_constant}, we display the LF
of sources with $a < 220\arcsec$ and a $<90\%$ probability of being
variable.
We fit
this LF (for 68 sources in the Analysis Sample with
$L_{\rm all} > 1.4 \times 10^{37} {\rm \, ergs \, s}^{-1}$) using the
same techniques we have used previously
\citep[\citetalias{SIB2000,SIB2001};][]{BSI2001,ISB2002}.
Since the average completeness correction factor in the Analysis
Sample
is only $\sim 1.01$, we do not apply completeness corrections
for this fit.
We adopted
the background LF from \cite{KWG+2004}; however, we have assumed that
background sources exhibit the same level of variability as the LMXBs
and reduced the expected background number from 11.7 to 8.2.

We modeled the LMXB populations with a single power law, a cutoff power
law, and a broken power law.
\begin{mathletters}
\begin{eqnarray}
  \label{eq:n4697x_lfs}
  {\rm Single}: \ \ \  
  \frac{ d N }{ d L_{37} } &=& 
     N_{0,s} \ 
       L_{37} ^ {-\alpha_s}; \\
  \label{eq:n4697x_lfc}
  {\rm Cutoff}: \ \ \ 
  \frac{ d N }{ d L_{37} } &=& 
     N_{0,c}
     \begin{cases}
       \left( \frac{L_X}{L_{\rm c}} \right)^{-\alpha_{\rm c}} & {\rm if \ } L_X \le L_{\rm c};\\
       0                                                      & {\rm otherwise};
     \end{cases}\\
  \label{eq:n4697x_lfb}
  {\rm Broken}: \ \ \ 
  \frac{ d N }{ d L_{37} } &=& 
     N_{0,b}
     \begin{cases}
       \left( \frac{L_X}{L_{\rm b}} \right)^{-\alpha_{\rm l}} & {\rm if \ } L_X \le L_{\rm b};\\
       \left( \frac{L_X}{L_{\rm b}} \right)^{-\alpha_{h}} & {\rm otherwise},\\
     \end{cases}
\end{eqnarray}
\end{mathletters}%
where $L_{37}$ is the X-ray luminosity in units of $10^{37} {\rm \,
ergs \, s}^{-1}$. We used the maximum
likelihood method to determine
the best fits to the cumulative LF and Monte Carlo techniques to
determine the errors (90\% confidence interval). A K-S test against
the cumulative distribution function of our best-fit LF indicated only
a 12\% chance that the single power law is a proper fit.
Much better fits were achieved for a cutoff power-law
($\Delta \chi^2 = \ $$-$10.6 for one less dof) with
$N_{0,c} = ( 9.5^{+22.5}_{-\phn5.6} ) \times 10^{-2}$,
$\alpha_{\rm c} = 1.48^{+0.21}_{-0.27}$, and
$L_{\rm c} = ( 6.0^{+3.8}_{-2.7} ) \times 10^{38} {\rm \, ergs \, s}^{-1}$
and for a broken power-law
($\Delta \chi^2 = \ $$-$14.3 for two less dof) with
$N_{0,b} = 2.2^{+3.0}_{-1.2}$,
$\alpha_{\rm l} = 1.02^{+0.30}_{-0.55}$,
$\alpha_{h} = 2.91^{+3.14}_{-0.59}$, and
$L_{\rm b} = ( 10.6^{+5.8}_{-4.4} ) \times 10^{37} {\rm \, ergs \, s}^{-1}$.
Although the broken power-law is the best fit according to
$\Delta \chi^2$, we note that the one-sided K-S test indicated the
cutoff power law model was acceptable (at the 50\% confidence level).
All three fits are overlaid in Figure~\ref{fig:n4697x_lf_constant}.

Since each of the five observations is an independent measure of the
instantaneous LF, there are enough datapoints that one can apply fits
of equations~\ref{eq:n4697x_lfs}--\ref{eq:n4697x_lfb} to a binned,
differential LF using standard $\chi^2$ techniques. In addition to the
five-fold increase in the number of datapoints, this analysis has the
advantage of including variable sources and producing a clearer
goodness-of-fit test than is provided by the K-S statistic. We chose
to combine the luminosities in bins of at least 25 instantaneous
luminosities.
Since there are five observations, each luminosity added 0.2 to the
instantaneous LF.
All fits were done for $L_X > 4\times10^{37} {\rm \, ergs \, s}^{-1}$
to avoid problems due to incompleteness and less accurate
measurements of $L_X$;
this resulted in 12 bins being used to fit the LFs.
For our best fit single, cutoff, and
broken power law LFs we found $\chi^2 = 25.1$, 14.0, and 4.9 for 10,
9, and 8 dof, which correspond to rejection probabilities of 99.5\%,
88\%, and 23\%. We believe that one of the reasons these rejection
probabilities are stronger than for the K-S statistic is that the K-S
test is less sensitive at the ends of its distribution. The single
power law is strongly rejected. The broken power-law is clearly the
best fit, with
$N_{0,b} = 3.1\pm1.5$,
$\alpha_{\rm l} = 0.83\pm0.52$,
$\alpha_{h} = 2.38\pm0.33$, and
$L_{\rm b} = ( 10.8\pm2.9 ) \times 10^{37} {\rm \, ergs \, s}^{-1}$;
however, we do not definitively rule out a cutoff
power law with
$N_{0,c} = 0.19 \pm 0.11$,
$\alpha_{\rm c} =1.55\pm0.18$, and
$L_{\rm c} = ( 4.9\pm1.1) \times10^{38} {\rm \, ergs \, s}^{-1}$
(90\% confidence intervals). We also checked for effects
due to errors in the background LF from \cite{KWG+2004} by determining
the best-fit LMXB LFs assuming 100 randomized realizations of the
background LF.
The single power law was rejected at the 98\% confidence level for
all realizations.
For 90\% of the realizations,
the cutoff power law can be rejected at better than the 85\%
confidence level (the lowest rejection is at the 70\% confidence
level). Including these effects would have little impact on our error
budget.

\begin{figure}
\plotone{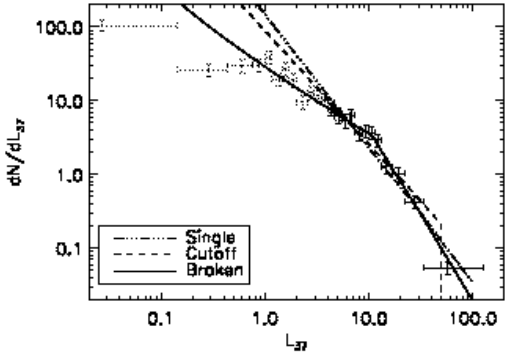}
\caption[Differential \, Instantaneous \, Luminosity \, Function \, of \, Discrete \, X-ray \, Sources \, in \, NGC~4697]
{
Completeness-corrected, instantaneous, differential luminosity
function from all five observations for sources detected within $a <
220\arcsec$.
Fits were performed on the solid data-points, whose
incompleteness correction is minimal.
The continuous curves are the sum
of the best-fit LMXB luminosity function and the expected
background source counts.
The broken power law is the best fit;
however, a cutoff power law cannot be rejected according to the
$\chi^2$ test (at the 85\% confidence level for a majority of fits
with different realizations of the background LF).
The reliability of both the best-fit broken power law LF
and the incompleteness correction factors are strengthened by the
broken power law LF going through many of the dotted data-points
(above $10^{37} {\rm \, ergs \, s}^{-1}$) that were not part of the fit.
\label{fig:n4697x_lf_inst}}
\end{figure}

We display the completeness-corrected, instantaneous, differential
luminosity function of the five observations for sources detected
within $a < 220\arcsec$ in Figure~\ref{fig:n4697x_lf_inst}. The
best-fit single, cutoff, and broken power law LFs derived above are
overlaid. The reliability of both the fitted LF and the independent
completeness correction factors (above $10^{37} {\rm \, ergs \, s}^{-1}$)
are strengthened by the broken power law LF going through many of the
low luminosity data-points that were not included in the fitting
process.

\begin{figure}
\plotone{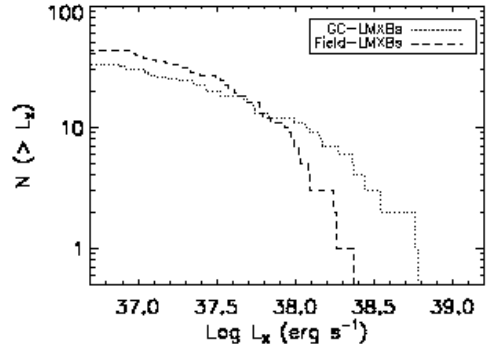}
\caption[Cumulative Luminosity Functions of Discrete X-ray Sources in NGC 4697: GC-LMXBs vs.\ Field-LMXBs]
{
Cumulative luminosity functions ($L_{X}=L_{\rm constant}$) of X-ray
sources in the HST-ACS FOV, excluding sources identified with non-GC
optical sources.
There are more bright X-ray sources in GCs;
however, this difference is not highly statistically significant.
\label{fig:n4697x_lf_gc_field}}
\end{figure}

In Figure~\ref{fig:n4697x_lf_gc_field} we have displayed the LFs
(using $L_{\rm all}$) of X-ray sources in the HST-ACS FOV that are in GCs
(34 GC-LMXBs) and in the field (44 Field-LMXBs).
After proper
renormalization, the distributions track each other very well when
$L_X < 3\times10^{37} {\rm \, ergs \, s}^{-1}$.
Above $6\times10^{37} {\rm \, ergs \, s}^{-1}$, there are always more
GC-LMXBs than Field-LMXBs.
A similar result was found by \citet{ALM2001} in NGC 1399, but was
not seen clearly in other samples of early-type galaxies
\citep{KMZ2002,SKI+2003,JCF+2004}.
Neither the K-S ($P_{\rm KS} = 0.33$) nor the Wilcoxon rank-sum test
$\sigma_{\rm WRS}=0.8$ indicate that the two LFs are drawn from a
different distribution. On the other hand, we can construct $2\times2$
contingency tables, comparing the numbers of Field-LMXBs and GC-LMXBs
below and above a given luminosity, and calculate Fisher's Exact Test
probabilities ($P_{\rm FE}$; \citealt{F1922}) that indicate $<$10\%
chance that the rows and columns are independent. Since probabilities
of independence calculated from contingency tables do not take into
account the freedom to choose the luminosity used to divide the
populations, the luminosity chosen must be physically motivated. If
we choose the luminosity corresponding to the Eddington limit for a
hydrogen accreting $1.4 {\rm \, M_\odot}$ NS ($1.8\times10^{38} {\rm
\, ergs \, s}^{-1}$), we find $P_{\rm FE}=0.036$.
While this is suggestive of a discrepancy between the LFs,
we do not believe the current data clearly indicates a statistically
significant difference between LFs of Field-LMXBs and GC-LMXBs.

\begin{figure}
\plotone{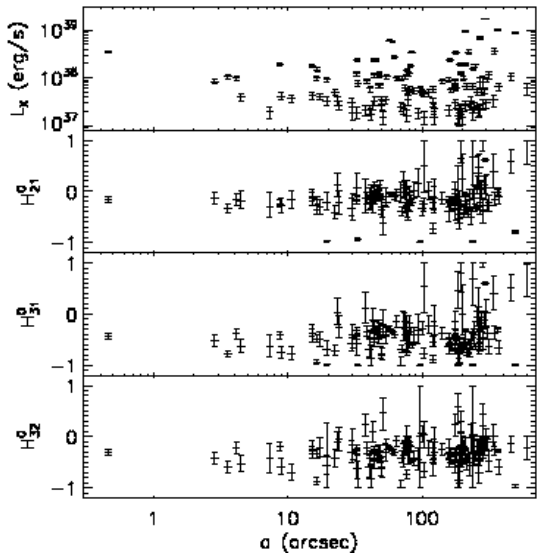}
\caption[Luminosity and Hardness Ratios Spatial Functions of Discrete X-ray Sources in NGC~4697]{
Merged luminosities ($L_{X}=L_{\rm constant}$) and hardness ratios as a
function of
the projected galactocentric semimajor distance, $a$,
for Analysis Sample sources.
No significant spatial difference in the luminosity function is observed.
Correlations
between hardness ratios and position (harder at greater distance)
appear to be associated with sources beyond 220$\arcsec$, and are likely due
to hard background AGNs as opposed to LMXBs in NGC~4697.
\label{fig:n4697x_lumhij_a}}
\end{figure}

We also searched for any variation of the LF with projected
galactocentric distance, using the Spearman's rank correlation coefficient
\citep[hereafter Spearman's $\rho$,][]{S1904},
which is a non-parametric test for correlations between two
properties.
Using both instantaneous and constant luminosities
for sources within $a<220\arcsec$, as well as constant luminosities over
the entire FOV, we found no significant evidence of a correlation
between luminosity and spatial position, $a$, for Analysis Sample
sources.
Similarly, no spatial difference in the constant luminosity
of significantly detected sources was indicated by a Wilcoxon rank-sum
comparison of two spatial bins, $a < a_{\rm eff}$ and $a > a_{\rm eff}$.
We display the luminosities of Analysis Sample sources
versus galactocentric semimajor axis in the top frame of
Figure~\ref{fig:n4697x_lumhij_a}.

\section{Hardness Ratios}
\label{sec:n4697x_src_colors}

The spectral properties of sources can be crudely characterized by
hardness ratios or X-ray colors (e.g., \citetalias{SIB2000,SIB2001}).
We defined hardness ratios of
$H_{21} \equiv (M - S)/(M + S)$,
$H_{31} \equiv (H - S)/(H + S)$,
and
$H_{32} \equiv (H - M)/(H + M)$,
where $S$, $M$, and $H$ are the total counts in the
soft (0.3--1 keV),
medium (1--2 keV),
and hard (2--6 keV) bands
\citep{SSC2004}.
Hardness ratios without superscripts are the measured values;
we use the superscript 0 to indicate the intrinsic hardness ratios,
correcting for Galactic absorption and QE degradation in the Chandra
ACIS detectors.
The counts in each band are corrected assuming the 
best-fit Chandra X-ray spectrum of the inner resolved
sources ($a < a_{\rm eff}$: Table~\ref{tab:spectra_n4697}, row 3).
Since the different observations had different QE
degradations, we adopted the following technique for correcting the
observed counts in a band.
Let $N_{i,j,k}$ be the net counts
for source $i$ in band $j$ during observation $k$,
and let $C_{j,k}$ be the
absorption and degradation correction to counts for observation $k$ in
band $j$.
(The QE degradation was assumed to be independent of position on the
detector and thus the same for all sources within a given observation.)
The combined correction for source $i$ in band $j$
is given by
$\langle C_{i,j} \rangle \equiv \underset{k}\sum (C_{j,k} N_{i,j,k}) /
                                \underset{k}\sum N_{i,j,k}$,
where the sums are only performed over observations where
$N_{i,j,k}>0$. We require that $N_{i,j,k}>0$ because there is no
correction to the source hardness when a source is not emitting at a
detectable level. Having determined the appropriate correction, the
corrected number of counts for source $i$ in band $j$ is $N^0_{i,j} =
\langle C_{i,j} \rangle \underset{k}\sum N_{i,j,k}$, where the sum is
now over all observations. The corrected band counts are then used to
calculate the corrected hardness ratios. We also combined Monte Carlo
simulations of the observed counts in the source and background
apertures with the count corrections to calculate the $1 \sigma$
confidence intervals for the hardness ratios. (When no counts are
observed, we set the expected number of counts in the Monte Carlo
simulations to 0.653. This is the average expectation for the Poisson
distribution for expected numbers of counts between 0 and 1.841, which
are the $1 \sigma$ confidence intervals on a measurement of zero
counts.)

\begin{figure*}
\epsfig{file=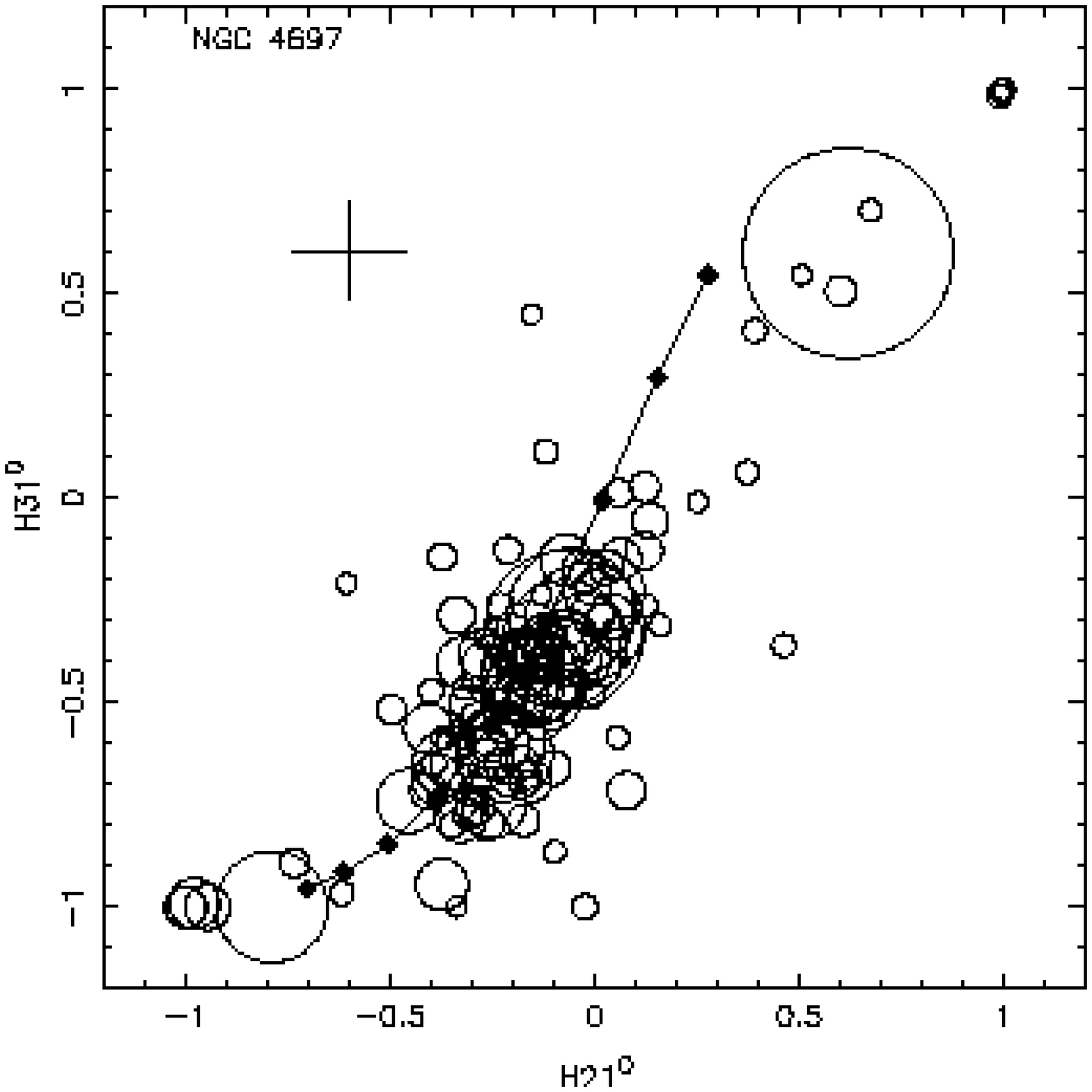, angle=-90, width=0.45\textwidth}
\hfil
\epsfig{file=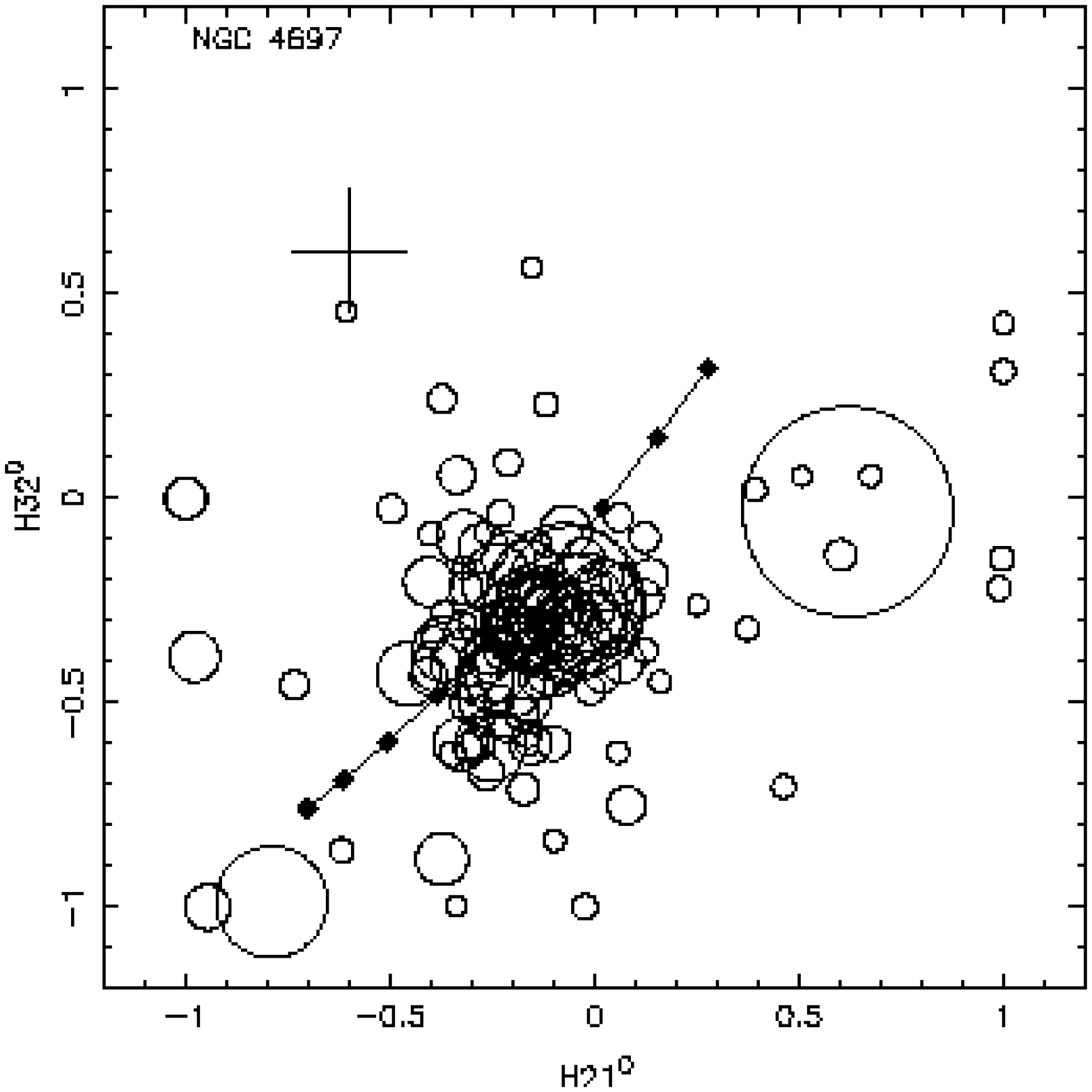, angle=-90, width=0.45\textwidth}
\caption[Hardness Ratio Diagram of Discrete X-ray Sources in NGC~4697]{
X-ray color-color diagrams 
$H_{31}^0$ vs.\ $H_{21}^0$ (left)
and
$H_{32}^0$ vs.\ $H_{21}^0$ (right)
for the combined counts from all observations
for the Analysis Sample sources.
Here, $H_{21}^0 \equiv (M^0 - S^0)/(M^0 + S^0)$, $H_{31}^0
\equiv (H^0 - S^0)/(H^0 + S^0)$, and $H_{32}^0 \equiv (H^0 - M^0)/(H^0
+ M^0)$, where $S^0$, $M^0$, and $H^0$ are the counts in the soft
(0.3--1 keV), medium (1--2 keV), and hard (2--6 keV) bands, corrected
for the effect of Galactic absorption and QE degradation according to
the best-fit spectra of resolved sources. The area of each circle is
proportional to the observed number of net counts. The solid curve and
large diamonds show the hardness ratios for power-law spectral models;
the diamonds indicate values of the power-law photon number index from
$\Gamma = 0$ (upper right) to 3.2 (lower left) in increments of
0.4. The model underwent the same correction as the sources. The
$1 \sigma$ error bars at the upper left illustrate the median of the
uncertainties.
\label{fig:n4697x_hij}}
\end{figure*}

X-ray color-color diagrams of the combined intrinsic hardness ratios
of the Analysis Sample sources are shown in Figure~\ref{fig:n4697x_hij}.
The values of the hardness ratios 
and their $1 \sigma$ errors are listed in
columns (5)--(7) of Table~\ref{tab:n4697x_lumcolor}.
Harder sources tend to lie in the upper
right of Figure~\ref{fig:n4697x_hij}a. Extra absorption tends to push
objects to the right in Figure~\ref{fig:n4697x_hij}b.

The majority of sources have hardness ratios consistent with power-law
indices of 1.2--2.0. On average, these sources tend to lie to the
right of the power-law curve, which might indicate some extra
absorption is occurring. A few sources occupy very different positions
in the hardness ratio planes. Sources 81, 96, 105, 112, 123, 134, 146,
148, 153, and 158 are harder than the typical source in NGC~4697. Both
their spectral properties and tendency to occur farther away from the
center of NGC~4697 suggest that these very hard sources may be
unrelated, strongly absorbed AGNs. Sources 11, 19, 25, 78, 84, 94,
110, 121, and 156 are softer than the typical source in
NGC~4697. Sources 84 and 156 both have optical counterparts; the
former is an extended object much redder than a typical GC, while the
latter is a star. Sources 19, 25, 78, and 110 have little if any emission
above 1 keV and are all SSs. Based on its $H_{32}$ color, Source 150
may be strongly absorbed or have a very atypical spectrum.

As with luminosity, we used the Spearman's $\rho$ test to search for a
correlation between merged hardness ratio and galactocentric semimajor
distance, $a$. We display the hardness ratios of Analysis Sample
sources versus spatial position in the bottom frames of
Figure~\ref{fig:n4697x_lumhij_a}. If we use the entire Analysis
Sample, we find $1.8 \sigma$, $2.2 \sigma$, and $1.8 \sigma$
significant correlations of $H_{21}^0$, $H_{31}^0$, and $H_{32}^0$,
respectively, with distance. In each case, the sense of the
correlation is that harder sources are found at larger $a$. These
correlation become insignificant ($<1 \sigma$) when only sources with
$a<220\arcsec$ are considered. The Wilcoxon rank-sum comparison of
$a<220\arcsec$ and $a>220\arcsec$ significantly detected sources
reproduces the effects seen by the Spearman's $\rho$ test. However,
we believe that the correlation at larger distances is due to the
increasing dominance of hard background AGNs, as opposed to LMXBs
intrinsic to NGC~4697. Scaling the expected number of sources
unrelated to NGC~4697 with the area in each region, we expect $\sim$12/97
and $\sim$17/29 Analysis Sample sources are unrelated to NGC~4697 for
$a<220\arcsec$ and $a>220\arcsec$, respectively.

\section{Spectral Analysis}
\label{sec:n4697x_src_spectra}

We performed an analysis of the spectra of sources in the $0.5$--$10.0 {\rm \,
keV}$ band, extracting the spectra and response files separately for
each observation of each source. The background spectra for each source
were determined locally, using the same nearby regions as discussed in
\S~\ref{sec:n4697x_detections}. Note that the response files for each
separate observation and source include the varying effects of
absorption by the contaminant which produces the QE degradation in the
ACIS detectors. Since the majority of our sources are too faint for
spectral analysis, we co-added the spectra and responses of groups of
sources for each observation. We only included sources whose total
count rate was determined at the $3\sigma$ level. We also excluded the
sources discussed individually in \S~\ref{sec:n4697x_ind_src_spectra}
and Source 110 (see Paper V). All of the spectra were grouped to
have at least 25 counts per spectral bin prior to background
correction to enable our use of $\chi^2$ statistics.
The use of a minimum number of counts per spectral bin and the restricted
energy range in the spectrum can result in some excluded bins, although
those bins have some photons in the allowed energy range.
Unless otherwise noted, the individual
observations were required to have the same spectral shape, but their
normalizations were allowed to vary.

The spectra of Galactic LMXBs can be complex. High luminosity LMXBs,
like those seen in NGC~4697, have often been modeled with
multi-component models that may include combinations of an isothermal
blackbody, a multi-color disk blackbody, a Comptonized power-law, or
more complicated models \citep{WNP1995}.
For our observations, there are two complicating factors.
First, we are
looking at a collection of sources, each of which is likely to have a
somewhat different spectrum.
Second, our spectra, even when co-added, have a much lower signal-to-noise
ratio than that for a Galactic LMXB.
Therefore, we have only
attempted to fit simple power-law or bremsstrahlung models to our
spectra. We summarize the results of the spectral fits in
Table~\ref{tab:spectra_n4697}. In each
row, we list how the sources were grouped together in the second
column. In the third column we list the model we used to fit the spectra.
In addition to the input model, we accounted for
an absorption column ($N_H$) using the Tuebingen-Boulder
absorption ({\scshape tbabs}) model assuming abundances from
\citet{WAM2000} and photoelectric absorption cross-sections from
\citet{VFK+1996}.
The absorbing column will typically be fixed at the Galactic
absorption column
\citep[$2.14 \times 10^{20} {\rm \, cm}^2$;][]{DL1990};
however, some fits require non-Galactic absorption. Absorption columns below the
Galactic value may indicate extra emission beyond the chosen model at low
energies, while absorption columns above the Galactic value may indicate the
presence of a local absorber or that the intrinsic spectra at low energies is
softer than the model spectra. The value of the absorbing column density ($N_H$)
is given in the fourth column. In the fifth column, we list either the
temperature $kT$ (for bremsstrahlung) or photon number spectral index $\Gamma$
(for a power-law). The sixth through tenth column list the unabsorbed fluxes $F$
(0.3--10 keV) of observations 0784, 4727, 4728, 4729, and 4730 respectively. The
last two columns give the total number of net counts in each set of spectra and
$\chi^2$ per dof for the best-fit model. All errors reported in the spectral
analysis are 90\% confidence level errors. Parentheses are used to indicate a
frozen parameter, and square brackets are used when an error is unconstrained on
one side.

\subsection{Best-Fit Spectra of LMXBs}

\begin{figure}
\hfil
\epsfig{file=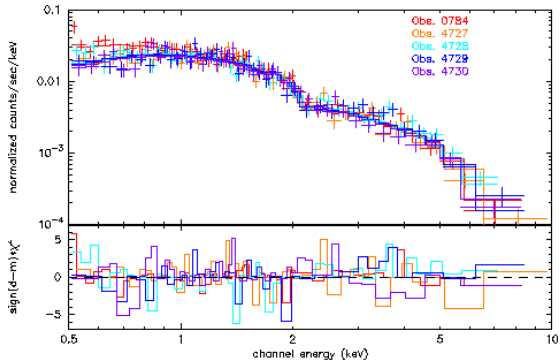, angle=-90, width=0.85\linewidth}
\hfil
\caption[X-ray Spectrum of LMXBs in NGC~4697]{
{\it Top panel}: Cumulative X-ray spectra of the resolved
sources in the inner effective elliptical isophote of NGC~4697
overlayed by the solid
histograms of the best-fit model spectra (Table~\ref{tab:spectra_n4697},
row 3).
The different observations are shown in different colors.
{\it Bottom  panel}: Contribution to
$\chi^2$ with the sign indicating the sign of the residual.
The differences between Observation 0784 and the Cycle-5 observations
at low energies are the result of increased QE degradation with time
due to a contaminant.
\label{fig:n4697x_bf_spec}}
\end{figure}

To determine the best-fit spectra for all LMXBs, we considered the
spectrum of all significantly detected sources within the elliptical
isophote that contains half of the optical light ($a < 1 a_{\rm eff}$;
Table~\ref{tab:spectra_n4697}, rows 1--4). Of the four fits to this
spectrum, those with bremsstrahlung models (bremss, rows 3 and 4) were
significantly better than those with power-law models (power, rows 1
and 2). For one dof, the $\Delta \chi^2 >12.8$. We note that the
derived power-law photon indices, $\Gamma = 1.47$--$1.66$ are
consistent with $\Gamma = 1.56 \pm 0.02$ found by \citet{IAB2003} for
LMXBs in a sample of early-type galaxies. The better fit to a
bremsstrahlung model compared to a power-law model has important
implications for estimating the unabsorbed flux of a source; the flux
conversion for the Galactic-absorbed bremsstrahlung model is $\sim
10\%$ lower than that for the Galactic-absorbed power-law model.
There is no evidence for a non-Galactic absorbing column in the
bremsstrahlung models ($\Delta \chi^2 = 0.08$ between row 3 and 4). We
adopt row 3 ($N_{H} = 2.14 \times 10^{20} {\rm \, cm}^2$, $kT =
9.1^{+1.3}_{-1.1} {\rm \, keV}$) as our best-fit spectra. We believe
that calibration changes may account for our spectra being slightly
harder than that found by \citet{IAB2003} ($kT= 7.3\pm0.3 {\rm \,
keV})$.
We display the observed spectra overlaid by the best-fit spectral model in
Figure~\ref{fig:n4697x_bf_spec}. Hereafter, we present results for a
bremsstrahlung model with Galactic absorption in different selections
of sources. We also present other models (power-law or non-Galactic
absorption) when they are statistically better fits.

\subsection{Spectra Grouped by Position}

In Table~\ref{tab:spectra_n4697}, rows 3, 4--6, 8 and 9, we compare
bremsstrahlung models with Galactic absorption for different collections of
sources grouped by position. These results mirror those presented previously
based on hardness ratios. When we compare the independent fits of
sources within $a < a_{\rm eff}$ (row 3) and in the annulus
$a_{\rm eff} < a < 2 a_{\rm eff}$ (row 5)
to the fit when their temperatures are tied
together (row 8), we find a $\Delta \chi^2 = 0.5$ for 1 dof. There is
no evidence for any spectral variation with galactocentric distance
within $2 a_{\rm eff}$.

On the other hand, there is evidence for a spectral change for sources within $2
a_{\rm eff} < a < 3 a_{\rm eff}$. First, comparing the fit to the spectra when
the temperatures within $3 a_{\rm eff}$ are grouped together (row 9) to the fits
when the temperatures can vary (rows 3, 5, and 6) indicates a slightly
significant statistical difference ($\Delta \chi^2 = 7.0$ for 2 dof and the
f-test indicates the probability that the free temperature model comes from the
tied temperature model is 4.0\%). Furthermore, a power-law fit (row 7) for $2
a_{\rm eff} < a < 3 a_{\rm eff}$ sources is a statistically better fit ($\Delta
\chi^2 = 4.3$ for 0 dof) than a bremsstrahlung fit (row 6). The increasing
dominance of background AGN in the spectra of regions further away from the
galaxy center is a likely cause for the apparent spatial variation of the
spectra. Our flanking-field HST observations will allow us to better address
this issue by better identifying background AGN.

\subsection{Spectra Grouped by GC Association}

We compare bremsstrahlung models in Table~\ref{tab:spectra_n4697} rows 10--18
for different collections of sources grouped by their association with
GCs in the HST-ACS FOV. Our comparisons of GC-LMXBs, Field-LMXBs, and
their combination (rows 10, 12, and 14; rows 11, 13, and 15) indicate
that the GC-LMXBs and Field-LMXBs are likely to be fit by different
spectrum. When a Galactic absorbing column is used, the $\Delta
\chi^2 = 8.6$ for 1 dof and the probability both populations have the
same temperature and absorption is $3.1\times10^{-3}$. The disparity is
even larger if we allow a non-Galactic absorbing column;
$\Delta \chi^2 = 14.0$ for 2 dof and the probability they have the
same temperature and absorption is $7.7\times10^{-4}$.

The best overall fit comes from allowing both GC-LMXBs and Field-LMXBs
to have independent absorbing columns and temperatures (rows 11 and 13).
The GC-LMXBs are better fit with a harder spectrum and larger
absorbing column compared to the Field-LMXBs. Although a larger
absorbing column in GC-LMXBs might explain this discrepancy, we note
that the absorption column of Field-LMXBs tends be pushed towards
sub-Galactic columns.
This suggests that the bremsstrahlung model underpredicts the
low-energy end of their spectra. A more accurate model and better
understanding of low energy calibration issues is required to better
probe the cause of the discrepancy in spectra between GC-LMXBs and
Field-LMXBs.

In \citet{MKZ2003}, the summed spectra of LMXBs in blue-GCs are found to be
harder than the summed spectra for LMXBs in red-GCs. This appears to be best
explained by an additional absorbing column in blue-GCs and is attributed to
irradiation induced winds. While extra absorption due to winds might explain why
our GC-LMXBs have a larger column than Field-LMXBs, we note that our data is not
entirely consistent with this scenario. When we compare the fits of spectra of
red-GC-LMXBs and blue-GC-LMXBs where they share the same absorbing column and
temperature to fits where they have independent absorbing columns and
temperatures, we find the $\Delta \chi^2 = 5.6$ for 2 dof and the probability
they have the same temperature and absorption is 6.1\%. These differences are
only marginal at best. Although the blue-GC LMXBs have a harder temperature than
red-GC LMXBs, they also have a lower absorbing column that tends towards being
sub-Galactic. Our understanding of the spectra of the blue-GC-LMXBs is limited
by their relatively small numbers. In addition to a more accurate model and
better understanding of low energy calibration issues, the spectra of
blue-GC-LMXBs from many galaxies should be combined to improve our statistics.

\subsection{Spectra Grouped by Luminosity}

In our final grouped spectral fits, we examine the different spectra
in two luminosity bins, $L_{38} < 2$ and $L_{38} > 2$, where
$L_{38} = L_X/10^{38} {\rm \, ergs \, s}^{-1}$.
Since the brighter sources are above the Eddington limit for a
hydrogen accreting $1.4 {\rm \, M_\odot}$ NS,
they are likely to be either
super-Eddington accreting NS-LMXBs or BH-LMXBs.
Since we believe AGN
may be significantly contaminating the spectra outside of $2 a_{\rm eff}$,
we restrict our sample to LMXBs interior to that semi-major distance.
We note that the brightest X-ray source
in this spatial region (Source 58) has $L_{38} \approx 6$.

Comparing the fainter LMXBs, the brighter LMXBs, and their combination (rows 19,
21, and 23; rows 20, 22, and 24) indicates that the fainter and brighter LMXBs
are likely to have different spectra. Separate spectra improve the fit by
$\Delta \chi^2 = 7.7$, and the probability both populations have the same
temperature and Galactic absorption is $4.4\times10^{-3}$. Again, the disparity
is larger if the groups are allowed to have different absorbing columns; the
$\Delta \chi^2 = 14.2$ for 2 dof and the probability both populations have the
same temperature and absorption is $5.5\times10^{-4}$. The best-fit comes from
allowing faint LMXBs and bright LMXBs to have different temperatures and
absorbing columns. Both populations tend to have the same temperature ($\sim 8
{\rm \, keV}$). The fainter LMXBs tend to have smaller (mostly sub-Galactic)
absorbing columns, while the brighter LMXBs tend to have a small excess
absorption column ($\Delta N_H = 3.8^{+2.9}_{-2.8} \times 10^{20} {\rm \,
cm}^2$), perhaps of local origin. If some fraction of the mass transferred
to the compact object is not accreted and ends up acting as a
source of absorption, the larger rate of accretion in brighter
sources might account for the extra absorption column.

\subsection{Individual Source Spectra}
\label{sec:n4697x_ind_src_spectra}

We can also explore the individual spectra of several sources. Although we
mainly consider sources with $L_{\rm all} > 6 \times 10^{38} {\rm \, ergs \,
s}^{-1}$, we also discuss the spectrum of the central source (Source 1). The
spectra were extracted in the identical manner as the grouped spectra above. For
these fits, the errors in fluxes and luminosities are only the scaled errors in
the count rates from the spectral fitting process. The spectrum of the variable
transient Source 110 is discussed in detail in Paper V.

{\it Source 1:} The central source in NGC~4697 could be a central AGN, an LMXB,
or a collection of confused LMXBs. Although the source is not particularly
bright (spectral fits made to 343.9 total net counts), we extracted its
individual spectra. The source is best fit by a power-law $\Gamma =
1.40^{+0.23}_{-0.22}$ with Galactic absorption ($\chi^2 = 11.6$ for 8 dof). The
weighted unabsorbed luminosity at NGC~4697 for this spectra, $(4.00\pm0.20)
\times 10^{38} {\rm \, ergs \, s}^{-1}$, is 12\% higher than that derived from
the best-fit spectral shape for sources within $a< a_{\rm eff}$, which is
consistent with the difference between power-law and bremsstrahlung models. As
this spectral fit is completely consistent with either a central AGN having a
low absorbing column or an LMXB (or several LMXBs), the nature of Source 1 is
still unknown.

{\it Source 117:} This source is known to be an AGN at $z=0.696$
(\citetalias{SIB2001}).
The total net counts fit by spectra for this source was 993.3, 
but there was a large variance between observations. (We fit to
49.3 net counts in Observation 4729 and 280.0 in Observation 4730.)
As we discuss in Paper V, the spectral
state
of this source in Observation 4729 appears to differ from that in the other
observations. We accounted for this by allowing Observation 4729 to
have a different absorption than the other observations. The best-fit
spectra ($\chi^2 = 30.1$ for 31 dof) involved absorbed power-law
($\Gamma = 1.55^{+0.24}_{-0.23}$) models. Observation 4729 had a much
larger absorbing column ($N_{H} = 3.4^{+1.9}_{-1.6} \times 10^{22}
{\rm \, cm}^2$) than the other observations ($N_{H} =
2.3^{+1.1}_{-1.0} \times 10^{21} {\rm \, cm}^2$). This fits implies
absorbed 0.5--$8.0 {\rm \, keV}$ X-ray luminosities of
$(%
1.18\pm0.08,
1.48\pm0.09,
1.00\pm0.08,
1.12\pm0.16, {\rm \ and \ }
1.43\pm0.09%
) \times 10^{44} {\rm \, ergs \, s}^{-1}$,
respectively, for Observations 0784, and 4727--4730.

{\it Source 134:} Source 134 is the brightest X-ray source in our observations
and has 1973.3 net counts in its spectrum. Its hardness ratios,
which vary between observations (see Paper V)
indicate it is harder than a typical LMXB. Its spectra are best-fit
($\chi^2 = 56.4$ for 65 dof) by a power-law model
($\Gamma=1.79^{+0.18}_{-0.18}$). Each observation has a different
absorbing column:
$(%
1.35^{+0.30}_{-0.27},
1.59^{+0.33}_{-0.30},
0.85^{+0.30}_{-0.27},
1.10^{+0.30}_{-0.26},$  and $
1.30^{+0.38}_{-0.35}
) \, 
\times \, 10^{22} {\rm \, cm}^2$,
respectively, for Observations 0784, and 4727-4730.
If Source 134 is at the distance of NGC~4697, its unabsorbed
0.3--$10.0 {\rm \, keV}$ X-ray luminosities are then
$(%
4.21\pm0.20,
4.92\pm0.23,
3.25\pm0.17,
3.77\pm0.21, {\rm \ and \ }
4.42\pm0.22%
) \times 10^{39} {\rm \, ergs \, s}^{-1}$.
In this case, Source 134 would be a heavily absorbed ULX. However, we
believe it is more likely that this source is a background AGN with
absorbed 0.5--$8.0 {\rm \, keV}$ X-ray fluxes of
$(%
1.36\pm0.07,
1.54\pm0.07,
1.16\pm0.06,
1.28\pm0.07, {\rm \ and \ }
1.44\pm0.07
) \times 10^{-13} {\rm \, ergs \, cm}^{-2} {\rm \,s}^{-1}$.

{\it Source 143:} This source appears to have an uncatalogued DSS
counterpart. Although the X-ray source is bright, Source 143 only has
enough counts for spectral fitting in Observation 4730 (150.4 net
counts). The best-fit model ($\chi^2 = 4.1$ for 4 dof) is a
Galactic-absorbed power-law ($\Gamma=1.92^{+0.35}_{-0.32}$). Its
absorbed 0.5--$8.0 {\rm \, keV}$ X-ray flux is $(2.59 \pm 0.21) \times
10^{-14} {\rm \, ergs \, cm}^{-2} {\rm \, s}^{-1}$. Its unabsorbed
0.3--$10.0 {\rm \, keV}$ X-ray luminosity at the distance of NGC~4697
is $(5.1 \pm 0.4) \times 10^{38} {\rm \, ergs \, s}^{-1}$ for this
spectral model.

{\it Source 155:} Like Source 143, Source 155 appears to have an uncatalogued
DSS counterpart. We fit spectra to 621.8 net X-ray counts from Observations
4727-4729. Although we tried power-law, bremsstrahlung, disk blackbody, and gas
models (apec), none of these spectra gave a good fit. The Galactic-absorbed
power-law ($\Gamma=1.26^{+0.12}_{-0.12}$) was the best fit we found ($\chi^2 =
34.5$ for 21 dof), with the model tending to underpredict emission below $\sim2
{\rm \, keV}$. This spectral model implies absorbed 0.5--$8.0 {\rm \, keV}$
X-ray fluxes of $(5.48 \pm 0.40, 6.17 \pm 0.44, 7.46 \pm 0.50)
\times 10^{-14} {\rm \, ergs \, cm}^{-2} {\rm \,s}^{-1}$,
respectively for Observations 4727-4729. The unabsorbed 0.3--$10.0
{\rm \, keV}$ X-ray luminosities at the distance of NGC~4697 are
$(1.07\pm0.08, 1.20\pm0.09, 1.45\pm0.10) \times 10^{39} {\rm \, ergs
\, s}^{-1}$, respectively.

{\it Source 156:} This source is clearly associated with the bright
foreground star \object{BD-05 3573}, whose optical colors are roughly
consistent with an early to middle G type star.
Its effective temperature is $T_{\rm eff} \sim 5300 {\rm \, K}$
\citep{G1999}.
Based on its optical magnitude, the distance to the star is likely
to be
$\sim 7$ -- $700 {\rm \, pc}$, for dwarf to giant luminosity classes,
respectively.
In the three observations during which it is in our FOV
(Observations 4727-4729), the spectrum contains 563.5 net counts.
Among likely, simple stellar models of X-ray emission, we
find the emission spectrum from collisionally-ionized diffuse gas
(apec) model that has $kT = 440^{+56}_{-47} {\rm \, eV}$ and a
heavy element abundance of $1.18 [>0.32]$ solar fits best.
The temperature, $\sim5\times10^{6} {\rm \, K}$,
is reasonable for that of an X-ray corona.
While the 0.3--$10.0 {\rm \, keV}$ X-ray fluxes are
 $(%
2.89\pm0.17,
1.68\pm0.15, {\rm \ and \ }
1.88\pm0.15,
) \times 10^{-14} {\rm \, ergs \, cm}^{-2} {\rm \,s}^{-1}$, the bolometric
flux is $\sim 60\%$ higher.
We note that the X-ray luminosities expected at the distance
of a giant G star are consistent with those of other K and G giant stars
with $T_{\rm eff} \sim 5300 {\rm \, K}$ \citep{G1999}.
One class of giant stars, FK Comae stars, are rapidly rotating
chromospherically active stars. Such activity might explain the X-ray
flaring (see Paper V) observed in this
source. Follow-up optical spectroscopy is necessary to determine the
spectral type and rotation speed of \object{BD-05 3573} and
to test whether it is an FK Comae star.

\section{Conclusions}
\label{sec:n4697x_conclusion}
Multi-epoch {\it Chandra} observations reveal a wealth of information
on LMXBs in NGC~4697, the nearest optically luminous elliptical galaxy.
We detect 158 sources, 126 of which have their count rates determined
at $\ge 3 \sigma$. Ten sources have optical counterparts in ground-based
catalogs, including a known AGN (Source 117) and the foreground star
\object{BD-05 3573} (Source 156). With our {\it Hubble} observations of the
galaxy center, we find 36 additional optical counterparts. Most importantly,
we identify 34 LMXBs clearly associated with GCs.

We confirm that GCs that are optically brighter ($4.5\sigma$) and redder
($3.0\sigma$) are more likely to contain GCs. We find that GCs with larger
encounter rates are also more likely to contain GCs ($5.5\sigma$). When we fit
the expected number ($\lambda_t$) of LMXBs in a GC, we find $\lambda_t
\propto \Gamma_{h}^{0.74^{+0.14}_{-0.13}} \,
(Z/Z_\odot)^{0.50^{+0.18}_{-0.16}}$, where $\Gamma_{h}$ is the encounter rate
and $Z/Z_\odot$ is the metallicity of the GC. Our results agree well with those
found for fainter X-ray sources in Galactic GCs \citep{PLA+2003} and LMXBs in
M87 \citep{JCF+2004}. These results are also consistent with
\citetalias{SJS+2007}; however, our NGC~4697 data set is included in that
analysis.

We detect sources with X-ray luminosities $> 6\times10^{36} {\rm
\, ergs \, s}^{-1}$. The fraction of LMXBs associated with GCs,
$f_{X,{\rm GC}}$, is $38.4^{+6.1}_{-5.7}\%$ and does not appear to
depend on X-ray luminosity. We find $10.7^{+2.1}_{-1.8}\%$ of GCs
contain an LMXB at the detection limit, although we note that it is
likely that the percentage of GCs with an active LMXB is even higher
due to the X-ray flux limit of the current observations. 
Furthermore,
our X-ray detections are not complete at the detection limit.
At the luminosity limit of our Analysis
Sample ($1.4\times10^{37} {\rm \, ergs \, s}^{-1}$), which is $>89\%$
complete, $8.1^{+1.9}_{-1.6}\%$ of GCs contain an LMXB.
This is the third deepest probe of the GC/LMXB connection in an
early-type galaxy to date. [At $3.4 {\rm \, Mpc}$, fainter
luminosities can be more easily reached with \object{Cen A}; however,
studies of the GC-LMXB connection in Cen~A are made more difficult
by its larger angular extent on the sky and recent star-formation
\citep{I1998}]. Deep observations of NGC~3379 probe deeper in luminosity;
however, only nine of its GCs contain LMXBs.]
At this same limit, there have been two
($1.3^{+1.7}_{-0.9}\%$) Galactic GCs containing LMXBs (NGC 6440 and NGC 6624)
over the history of X-ray astronomy. The discrepancy between the Milky Way and
NGC 4697 may be explained by their different GC metallicity distributions (a 3:1
metal-poor to metal rich ratio for Galactic GCs as compared to a 1:1 ratio for
NGC 4697 GCs). Since metal-rich GCs ($[{\rm Fe/H}] > -0.75$) are about 3 times as
likely to contain LMXBs, NGC 4697 is predicted to have about twice the
percentage of GCs with LMXBs as the Milky Way. This correction eliminates most
of the discrepancy between the two galaxies.

We have determined the X-ray luminosity functions from each individual
observation, from the combination of our five observations, and the LF of the
non-variable sources. There is no statistically significant difference in the
LFs of the different observations. This result is critical because it validates
using single-epoch observations to measure LFs. While we clearly rule out a
single power-law LF, we cannot definitively rule out cutoff power-law models
with slopes of $\alpha = 1.5\pm0.2$ and cutoff luminosities of
$(6^{+4}_{-3}) \times 10^{38} {\rm \, erg\,s}^{-1}$. Broken power-law
models (eq.~[\ref{eq:n4697x_lfb}]) provide the best fits to our LFs. We adopt
our fit of the instantaneous LF as our best-fit, with $N_{0,b} = 3.1\pm1.5$,
$\alpha_{\rm l} = 0.83\pm0.52$, $\alpha_{h} = 2.38\pm0.33$, and $L_{\rm b} = (
10.8\pm2.9 ) \times 10^{37} {\rm \, ergs \, s}^{-1}$. We note that \cite{KF2004}
found evidence for a possible break in the LF of LMXBs in early-type galaxies at
slightly larger luminosities; however recent deep observations of NGC~3379 and
NGC~4278 have not found strong evidence for such a break \citep{KFK+2006}. This
raises the possibility that there is no universal form for the LF of LMXBs in
early-type galaxies.

We find marginal evidence (significant at the $2.1\sigma$ level) that a larger
number of LMXBs above the Eddington limit for a hydrogen accreting $1.4 {\rm
\, M_\odot}$ NS tend to be found in GCs than in the field. Although this
is consistent with results in \citet{ALM2001} in NGC 1399, we believe
this result needs to be tested with a larger sample. One possible
\citep[e.g.,][]{KMZ2007}
explanation is that multiple LMXBs might exist in some GCs. We predict that this
effect is small, only $\sim4$ of the GCs are expected to have multiple LMXBs
with total X-ray luminosity above $1.4 \times 10^{37} {\rm \, ergs \, s}^{-1}$,
which corresponds to $\sim 16\%$ of the GCs above that X-ray luminosity.
An alternative explanation is that any possible discrepancy in the LFs occurs
at lower luminosities. Such discrepancies have been seen for the bulge of M31
and the Milky Way \citep{VG2007} and the elliptical galaxy NGC~3379
\citep{FBZ+2008}; however, this effect is most evident at luminosities below
those probed by our observations of NGC~4697.

Our spectrum of sources in the inner effective semi-major axis is best-fit by
bremsstrahlung emission with Galactic absorption ($kT = 9.1^{+1.3}_{-1.1} {\rm
\, keV}$, $N_{H} = 2.14 \times 10^{20} {\rm \, cm}^2$). Both hardness ratios and
spectral analysis indicate that the spectra of X-ray sources at large radii ($a
\gtrsim 180\arcsec$) differ from those at small radii. We believe that this
effect is due to an increasing dominance of unrelated foreground and background
sources, particularly background AGNs. The spectra of GC-LMXBs and Field-LMXBs
appear to differ (significant at the $3.3\sigma$ level). The GC-LMXBs are better
fit with higher temperatures and greater absorbing columns compared to the
Field-LMXBs. Similarly, we find a difference (significant at the $3.5\sigma$
level) between X-ray fainter and brighter LMXBs. The fainter LMXBs tend to have
smaller absorbing columns, while the brighter LMXBs tend to have a small excess
in absorption, which may be due to having more accreting material. We find that
spectra of LMXBs in metal-poor GCs have harder temperature and lower absorbing
columns than those in metal-rich GCs; however, this marginal result is only
significant at the $1.9\sigma$ level. In all cases, the sources with a smaller
absorption column tend to be fit with sub-Galactic absorbing columns. This
indicates that the spectral model (folded in with the calibration) underpredicts
the soft emission in the spectra. To probe the cause of the spectral
differences, we require a more physically accurate model and better
understanding of the calibration at low energies.

Among the spectral fits to individual bright sources,
Source 117 (a known AGN) and Source 134 had spectral fits with
large absorbing columns. We predict Source 134 is likely to be an AGN.
The spectral fit and variability in Source 156 (the foreground star
\object{BD-05 3573}) are consistent with it being an FK Comae star,
a chromospherically active giant.

\acknowledgements

We thank Peter Frinchaboy and Rachel Osten for very helpful
discussions. Support for this work was provided by NASA through {\it Chandra}
Award Numbers GO4-5093X, AR4-5008X, and GO5-6086X, and through HST Award Numbers
HST-GO-10003.01-A, HST-GO-10582.02-A, and HST-GO-10597.03-A. G.~R.~S.\
acknowledges the receipt of an ARCS fellowship and support provided by the F. H. 
Levinson Fund.

\bibliography{ms}


\clearpage
\LongTables
\tabletypesize{\scriptsize}

\begin{deluxetable}{lcccccccccl}
\setlength{\tabcolsep}{3.78pt}
\tablewidth{0pt}
\tablecaption{Discrete X-ray Sources in NGC~4697 \label{tab:n4697x_src}}
\tablehead{
Source& 
&
R.A.&
Dec.&
\multicolumn{1}{c}{$d$}&
\multicolumn{1}{c}{$a$}&
\multicolumn{1}{c}{Count Rate}&
&
&
Paper II&
\\
No.&
Name&
(h m s)&
($\arcdeg$ $\arcmin$ $\arcsec$)&
\multicolumn{1}{c}{($\arcsec$)}&
\multicolumn{1}{c}{($\arcsec$)}&
\multicolumn{1}{c}{($10^{-4}$ s$^{-1}$)}&
\multicolumn{1}{c}{S/N}&
$c_{\rm sim}$&
Source&
Notes \\
(1)&
(2)&
(3)&
(4)&
(5)&
(6)&
(7)&
(8)&
(9)&
(10)&
\multicolumn{1}{c}{(11)}
}
\startdata
1  &CXOU J124835.8$-$054802&12 48 35.86&$-$05 48 02.7&\phn\phn0.45&\phn\phn0.46&   \phn25.57$\pm$1.26&   20.24&1.00&1      &c\\
2  &CXOU J124836.0$-$054803&12 48 36.03&$-$05 48 03.3&\phn\phn2.26&\phn\phn2.84&\phn\phn5.58$\pm$0.63&\phn8.86&1.00&2      &a\\
3  &CXOU J124835.8$-$054759&12 48 35.81&$-$05 47 59.9&\phn\phn2.89&\phn\phn4.47&\phn\phn2.77$\pm$0.50&\phn5.57&1.00&\nodata&\nodata\\
4  &CXOU J124835.6$-$054804&12 48 35.68&$-$05 48 04.1&\phn\phn3.53&\phn\phn3.56&\phn\phn7.60$\pm$0.75&   10.16&1.00&3      &\nodata\\
5  &CXOU J124836.1$-$054802&12 48 36.15&$-$05 48 02.1&\phn\phn3.96&\phn\phn4.12&\phn\phn6.73$\pm$0.68&\phn9.91&1.00&4      &a\\
6  &CXOU J124835.6$-$054757&12 48 35.65&$-$05 47 57.8&\phn\phn5.88&\phn\phn8.87&\phn\phn2.95$\pm$0.50&\phn5.93&1.00&5      &\nodata\\
7  &CXOU J124835.9$-$054756&12 48 35.96&$-$05 47 56.5&\phn\phn6.07&\phn\phn8.69&   \phn13.34$\pm$0.93&   14.40&1.00&6      &\nodata\\
8  &CXOU J124836.2$-$054758&12 48 36.29&$-$05 47 58.7&\phn\phn7.07&\phn\phn7.29&\phn\phn1.39$\pm$0.40&\phn3.53&1.04&\nodata&b,f\\
\enddata
\tablecomments{The simulated completeness correction factor, $c_{\rm sim}$,
 is accurate to 5\% for the majority of sources.}
\tablenotetext{a}{Source may be confused with nearby sources, making its position,
flux, and extent uncertain.}
\tablenotetext{b}{Source region noticeably more extended than PSF.}
\tablenotetext{c}{Possible cataloged optical counterpart may be the central AGN.}
\tablenotetext{d}{Possible cataloged optical counterpart appears to be a globular cluster.}
\tablenotetext{e}{Possible cataloged optical counterpart is known to not be a globular cluster.}
\tablenotetext{f}{Possible cataloged optical counterpart does not appear to be a globular cluster.}
\tablenotetext{g}{Possible uncataloged optical counterpart on DSS.}
\tablenotetext{h}{Possible radio counterpart.}
\tablenotetext{i}{Source may exhibit variability.}
\setlength{\tabcolsep}{6pt}
\end{deluxetable}

\clearpage
\tabletypesize{\footnotesize}
\begin{landscape}
\begin{deluxetable}{llcll}
\setlength{\tabcolsep}{6pt}
\tablewidth{0pt}
\tablecaption{Optical Properties of Ground-Based Optical Sources Matched to X-ray Sources in NGC~4697 \label{tab:n4697xo_gb_src}}
\tablehead{
X-ray Source& 
& 
Offset&
&
\\
No.&
Optical Source
($\arcsec$)&
Photometry &
Optical Notes\\
\\
(1)&
(2)&
(3)&
(4)&
(5)}
\startdata
1   & Center of NGC 4697      & 0.5 &                                                                        & \\
8   & 2MASS J12483635-0547577 & 1.4 & $J=13.1$, $H>10.9$, $K_{\rm s}>10.7$                                   & Galaxy light may contaminate photometry and PSF-fitting\\
    &                         &     &                                                                        & Optically extended?\\
15  & 2MASS J12483692-0548007 & 0.6 & $J=14.4$, $H=13.7$, $K_{\rm s}>11.7$                                   & Galaxy light may contaminate photometry and PSF-fitting\\
    &                         &     &                                                                        & Optically extended?\\
55  & 2MASS J12483648-0548475 & 0.2 & $J=16.6$, $H=16.2$, $K_{\rm s}>14.5$                                   & Galaxy light may contaminate photometry\\
    &                         &     &                                                                        & Optical point source\\
84  & 2MASS J12483420-0549261 & 0.5 & $J=16.0$, $H=15.5$, $K_{\rm s}=15.1$                                   & Optical point source\\
\enddata
\setlength{\tabcolsep}{6pt}
\end{deluxetable}

\clearpage
\end{landscape}

\clearpage
\tabletypesize{\footnotesize}
\LongTables

\begin{deluxetable}{lccccccc}
\setlength{\tabcolsep}{6pt}
\tablewidth{0pt}
\tablecaption{Optical Properties of HST Sources Matched to X-ray Sources in NGC~4697 \label{tab:n4697xo_hst_src}}
\tablehead{
X-ray Source& 
Offset&
$r_{h}$&
$g_{475}$&
$z_{850}$&
$g-z$&
\\
No.&
($\arcsec$)&
(pc)&
(AB)&
(AB)&
(AB)&
Notes \\
(1)&
(2)&
(3)&
(4)&
(5)&
(6)&
(7)}
\startdata
7 &0.05&$3.24\pm\phn0.18$&$21.49\pm 0.05$&$20.43\pm 0.03$ &$1.06\pm 0.06$ &a\\
15&0.11&$2.90\pm\phn0.11$&$19.30\pm 0.04$&$17.95\pm 0.02$ &$1.35\pm 0.04$ &b\\
16&0.06&$1.95\pm\phn0.14$&$23.10\pm 0.05$&$21.64\pm 0.04$ &$1.47\pm 0.06$ &b\\
18&0.80&$2.02\pm\phn0.20$&$24.58\pm 0.07$&$23.18\pm 0.03$ &$1.40\pm 0.08$ &b\\
21&0.44&$2.84\pm\phn0.11$&$20.95\pm 0.04$&$19.68\pm 0.01$ &$1.27\pm 0.04$ &b\\
22&0.06&\nodata          &\nodata        &\nodata         &\nodata        &f\\
  &0.25&\nodata          &\nodata        &\nodata         &\nodata        &f\\
24&0.10&$1.95\pm\phn0.22$&$22.89\pm 0.04$&$21.56\pm 0.04$ &$1.33\pm 0.06$ &b\\
\enddata
\tablenotetext{a}{Optical counterpart is a blue GC.}
\tablenotetext{b}{Optical counterpart is a red GC.}
\tablenotetext{c}{Optical counterpart rejected as GC since $g_{475}-z_{850} < 0.5$.}
\tablenotetext{d}{Optical counterpart rejected as GC since $g_{475}-z_{850} > 1.9$.}
\tablenotetext{e}{Half-light radius is poorly constrained. Therefore, we have used
 aperture based photometry. The $(g-z)$ color suggest this source is not a GC.}
\tablenotetext{f}{Multiple nearby optical counterparts prevented {\sc kingphot} analysis.}
\setlength{\tabcolsep}{6pt}
\end{deluxetable}

\clearpage
\tabletypesize{\footnotesize}
\LongTables

\begin{deluxetable}{lcccccc}
\setlength{\tabcolsep}{6pt}
\tablewidth{0pt}
\tablecaption{Fits to the Expected Number of LMXBs per GC ($\lambda$)
\label{tab:n4697gclmxb_par}}
\tablehead{
\multicolumn{7}{c}{$\lambda = A \, M^{\alpha} \,
(\mathcal{Z}/\mathcal{Z}_\odot)^{\beta} \,
( r ) ^{\delta}
{\rm \ or \ }   \lambda = A \, \Gamma^{\epsilon} \,
(\mathcal{Z}/\mathcal{Z}_\odot)^{\beta}$}\\
\colhead{Row}&
\colhead{$\alpha$}&
\colhead{$\beta$}&
\colhead{$\delta$}&
\colhead{$\epsilon$}&
\colhead{$\psi$}&
\colhead{$\Delta$ dof}
}
\startdata
1                  & $[0]$                     & $[0]$                  & $[0]$                   & $[0]$                     & $-101.16   $ & 0\\
2                  & $0.99^{+0.21}_{-0.21}$    & $[0]$                  & $[0]$                   & $[0]$                     & $\phn-87.41$ & 1\\
3                  & $[0]$                     & $0.39^{+0.15}_{-0.14}$ & $[0]$                   & $[0]$                     & $\phn-95.97$ & 1\\
4\tablenotemark{a} & $[0]$                     & $[0]$                  & $-2.02^{+0.58}_{-0.58}$ & $[0]$                     & $\phn-94.62$ & 1\\
5\tablenotemark{b} & $[0]$                     & $[0]$                  & $-1.55^{+0.57}_{-0.59}$ & $[0]$                     & $\phn-97.27$ & 1\\
6\tablenotemark{a} & $[0]$                     & $[0]$                  & $[0]$                   & $0.89 ^{+0.18}_{-0.16}$   & $\phn-77.75$ & 1\\
7\tablenotemark{b} & $[0]$                     & $[0]$                  & $[0]$                   & $0.87 ^{+0.19}_{-0.16}$   & $\phn-79.95$ & 1\\
8\tablenotemark{a} & $1.28^{+0.29}_{-0.25}$    & $[0]$                  & $-3.35^{+0.79}_{-0.87}$ & $[0]$                     & $\phn-76.25$ & 2\\
\enddata
\tablenotetext{a}{Parameters calculated assuming $r=r_h$ and $\Gamma \propto M^{3/2}({r_h})^{-5/2}$.}
\tablenotetext{b}{Parameters calculated assuming $r=r_h^\prime$ and $\Gamma \propto M^{3/2}({r_h^\prime})^{-5/2}$.}
\setlength{\tabcolsep}{6pt}
\end{deluxetable}

\clearpage
\tabletypesize{\footnotesize}
\LongTables

\begin{deluxetable}{lrrrrr}
\tablewidth{0pt}
\tablecaption{Instantaneous Luminosities of X-ray Sources \label{tab:n4697x_luminst}}
\tablehead{
\colhead{Source}&
\colhead{$L_{\rm A}$}&
\colhead{$L_{\rm B}$}&
\colhead{$L_{\rm C}$}&
\colhead{$L_{\rm D}$}&
\colhead{$L_{\rm E}$}
\\
\colhead{(1)}&
\colhead{(2)}&
\colhead{(3)}&
\colhead{(4)}&
\colhead{(5)}&
\colhead{(6)}
}
\startdata
   1                  &   40.3$\pm$\phn4.1 &   32.6$\pm$4.0&   34.1$\pm$4.2 &   36.2$\pm$\phn4.6 &   35.3$\pm$4.0 \\
   2                  &    6.6$\pm$\phn2.0 &   10.7$\pm$2.6&    8.9$\pm$2.5 &   10.4$\pm$\phn2.8 &    5.8$\pm$2.0 \\
   3                  &    2.5$\pm$\phn1.6 &    3.6$\pm$1.7&    3.5$\pm$1.8 &    6.4$\pm$\phn2.3 &    3.9$\pm$1.8 \\
   4                  &   11.6$\pm$\phn2.5 &    7.4$\pm$2.2&   11.8$\pm$2.8 &    9.7$\pm$\phn2.8 &   12.5$\pm$2.6 \\
   5                  &    7.1$\pm$\phn2.1 &    9.4$\pm$2.4&   13.6$\pm$3.0 &   10.4$\pm$\phn2.8 &   10.6$\pm$2.5 \\
   6                  &    3.4$\pm$\phn1.5 &    5.9$\pm$2.0&    6.3$\pm$2.2 &    4.7$\pm$\phn2.1 &    0.9$\pm$1.2 \\
   7                  &   21.5$\pm$\phn3.1 &   22.3$\pm$3.3&   12.6$\pm$2.8 &   21.8$\pm$\phn3.7 &   15.0$\pm$2.8 \\
   8                  &    1.8$\pm$\phn1.4 &    2.1$\pm$1.6&    3.5$\pm$1.7 &    1.1$\pm$\phn1.7 &    1.2$\pm$1.3 \\
\enddata
\tablecomments{Individual luminosities (in units of $10^{37} {\rm \, ergs \,
s}^{-1}$) of Observations 0784, 4727, 4728, 4729, 4730 are indicated
by $L_{\rm A}$, $L_{\rm B}$, $L_{\rm C}$, $L_{\rm D}$, and $L_{\rm
E}$, respectively.}
\end{deluxetable}

\clearpage
\tabletypesize{\footnotesize}
\LongTables

\begin{deluxetable}{lccrrrr}
\tablewidth{0pt}
\tablecaption{Combined Luminosities \& Hardness Ratios of X-ray Sources
\label{tab:n4697x_lumcolor}}
\tablehead{
\colhead{Source}&
\colhead{Not Detected}&
\colhead{Not in FOV}&
\colhead{$L_{\rm all}$}&
\colhead{$H21_{\rm all}^{0}$}&
\colhead{$H31_{\rm all}^{0}$}&
\colhead{$H32_{\rm all}^{0}$}
\\
\colhead{(1)}&
\colhead{(2)}&
\colhead{(3)}&
\colhead{(4)}&
\colhead{(5)}&
\colhead{(6)}&
\colhead{(7)}
}
\startdata
   1                  & \cennodata & \cennodata &   35.8$\pm$1.8 & $-$$0.16^{+0.05}_{-0.05}$ & $-$$0.44^{+0.05}_{-0.05}$ & $-$$0.30^{+0.05}_{-0.06}$\\
   2                  & ABCDE      & \cennodata &    8.4$\pm$0.9 & $-$$0.13^{+0.12}_{-0.11}$ & $-$$0.53^{+0.11}_{-0.10}$ & $-$$0.43^{+0.12}_{-0.11}$\\
   3                  & ABCDE      & \cennodata &    3.9$\pm$0.7 & $-$$0.18^{+0.18}_{-0.16}$ & $-$$0.64^{+0.16}_{-0.12}$ & $-$$0.53^{+0.17}_{-0.16}$\\
   4                  &  B         & \cennodata &   10.6$\pm$1.0 & $-$$0.33^{+0.09}_{-0.08}$ & $-$$0.78^{+0.07}_{-0.06}$ & $-$$0.60^{+0.11}_{-0.10}$\\
   5                  & \cennodata & \cennodata &   10.1$\pm$1.0 & $-$$0.17^{+0.11}_{-0.10}$ & $-$$0.39^{+0.11}_{-0.10}$ & $-$$0.23^{+0.11}_{-0.11}$\\
   6                  & AE         & \cennodata &    4.1$\pm$0.7 & $-$$0.29^{+0.16}_{-0.14}$ & $-$$0.76^{+0.12}_{-0.10}$ & $-$$0.60^{+0.18}_{-0.15}$\\
   7                  & \cennodata & \cennodata &   18.7$\pm$1.3 & $-$$0.24^{+0.07}_{-0.07}$ & $-$$0.42^{+0.07}_{-0.07}$ & $-$$0.20^{+0.08}_{-0.08}$\\
   8                  & ABCDE      & \cennodata &    2.0$\pm$0.6 & $-$$0.31^{+0.29}_{-0.21}$ & $-$$0.65^{+0.24}_{-0.19}$ & $-$$0.43^{+0.29}_{-0.27}$\\
\enddata
\tablecomments{
Observations 0784, 4727, 4728, 4729, 4730 are indicated by A, B, C, D, and E,
respectively.
If an observation is listed in the ``Not in FOV'' column, it means the
source was not in the S3 FOV for that observation, and thus could not
possibly be detected.
If an observation is listed in the ``Not Detected'' column, it means the
source was in the FOV for that observation, but was not detected by
{\sc wavdetect} in that individual observation.
Sources with ABCDE in the ``Not Detected'' were not detected in any
individual observation, only in the combined sum of the observations.
Luminosities are in units of $10^{37} {\rm \, ergs \,
s}^{-1}$.
}
\end{deluxetable}

\clearpage
\tabletypesize{\footnotesize}
\begin{landscape}
\LongTables

\begin{deluxetable}{lccccccccccrr}
\tablewidth{0pt}
\tablecaption{X-ray Spectral Fits of Point Sources in NGC~4697 \label{tab:spectra_n4697}}
\tablehead{
&
&
&
\colhead{$N_H$}&
\colhead{$kT$ or $\Gamma$}\\
\colhead{Row}&
\colhead{Group\tablenotemark{a}}&
\colhead{Model}&
\colhead{($10^{20}$ cm$^{-2}$)}&
\colhead{(keV)}&
\colhead{$F_{0784}$\tablenotemark{b}}&
\colhead{$F_{4727}$\tablenotemark{b}}&
\colhead{$F_{4728}$\tablenotemark{b}}&
\colhead{$F_{4729}$\tablenotemark{b}}&
\colhead{$F_{4730}$\tablenotemark{b}}&
\colhead{Counts}&
\colhead{$\chi^2$/dof}\\
\colhead{(1)}&
\colhead{(2)}&
\colhead{(3)}&
\colhead{(4)}&
\colhead{(5)}&
\colhead{(6)}&
\colhead{(7)}&
\colhead{(8)}&
\colhead{(9)}&
\colhead{(10)}&
\colhead{(11)}&
\colhead{(12)}
}
\startdata
1                  & $a < 1 a_{\rm eff}$                   & Power  & (2.14)              & $1.505^{+0.048}_{-0.032}$ & 3.77 & 3.71 & 3.87 & 3.79 & 3.59 &  7190.3 & 294.26/245=1.20\\
2                  & $a < 1 a_{\rm eff}$                   & Power  & $7.0^{+2.5}_{-2.4}$ & $1.605^{+0.061}_{-0.060}$ & 3.82 & 3.74 & 3.88 & 3.82 & 3.64 &  7190.3 & 283.16/244=1.16\\
3\tablenotemark{c} & $a < 1 a_{\rm eff}$                   & Bremss & (2.14)              & $9.1^{+1.3}_{-1.1}$       & 3.42 & 3.34 & 3.46 & 3.41 & 3.26 &  7190.3 & 270.49/245=1.10\\
4                  & $a < 1 a_{\rm eff}$                   & Bremss & $1.8^{+1.9}_{-1.8}$ & $9.3^{+1.9}_{-1.5}$       & 3.42 & 3.34 & 3.46 & 3.41 & 3.26 &  7190.3 & 270.41/244=1.11\\
5                  & $1 a_{\rm eff} < a < 2 a_{\rm eff}$   & Bremss & (2.14)              & $7.8^{+3.3}_{-2.5}$       & 0.82 & 0.84 & 0.82 & 0.90 & 0.78 &  1587.4 &   60.39/57=1.06\\
6                  & $2 a_{\rm eff} < a < 3 a_{\rm eff}$   & Bremss & (2.14)              & $5.7^{+1.6}_{-1.1}$       & 0.91 & 0.53 & 0.62 & 0.76 & 0.72 &  1652.9 &   68.07/64=1.06\\
7                  & $2 a_{\rm eff} < a < 3 a_{\rm eff}$   & Power  & (2.14)              & $1.649^{+0.079}_{-0.077}$ & 1.00 & 0.58 & 0.71 & 0.86 & 0.79 &  1652.9 &   63.80/64=1.00\\
8                  & $a < 2 a_{\rm eff}$                   & Bremss & (2.14)              & $8.9^{+1.2}_{-1.0}$       & 4.18 & 4.12 & 4.22 & 4.24 & 3.99 &  8777.7 & 331.35/303=1.09\\
\enddata
\tablenotetext{a}{We abbreviate $L_X/10^{38} {\rm \, erg \, s}^{-1}$ in the 0.3--10 keV band as $L_{38}$.}
\tablenotetext{b}{Units are $10^{-13} {\rm \, erg \, cm}^{-2} {\rm \,s}^{-1}$ in the 0.3--10 keV band.}
\tablenotetext{c}{The adopted best-fit model for LMXBs.}

\end{deluxetable}

\clearpage
\end{landscape}

\end{document}